\documentclass[fleqn,12pt,twoside]{article}
\usepackage[headings]{espcrc1}
\usepackage{epsfig}
\usepackage{lscape}
\readRCS
$Id: espcrc1.tex,v 1.2 2004/02/24 11:22:11 spepping Exp $
\title{Linear and Branched Polymers on Fractals}

\author{
Deepak Dhar\address{
Department of Theoretical Physics, \\ 
Tata Institute of Fundamental Research, \\ 
Homi Bhabha Road, Mumbai 400005, India}
and
Yashwant Singh \address
{Department of Physics,\\
Banaras Hindu University,\\
Varanasi, U.P. 221005, India.}}

\runtitle{Linear polymers on Fractals}
\runauthor{Singh and Dhar}

\begin{document}

\maketitle

\bigskip 
\begin{abstract} 

This is a pedagogical review of the subject of linear polymers on
deterministic finitely ramified fractals.  For these, one can determine
the critical properties exactly by real-space renormalization group
technique.  We show how this is used to determine the critical exponents
of self-avoiding walks on different fractals. The behavior of critical
exponents for the $n$-simplex lattice in the limit of large $n$ is
determined. We study self-avoiding walks when the fractal
dimension of the underlying lattice is just below $2$. We then consider
the case of linear polymers with attractive interactions, which on some
fractals leads to a collapse transition.  The fractals also provide a
setting where the adsorption of a linear chain near on attractive
substrate surface and the zipping-unzipping transition of two mutually
interacting chains can be studied analytically. We also discuss briefly
the critical properties of branched polymers on fractals.

\end{abstract}

\section{INTRODUCTION}

The problem of self-avoiding walks ( SAWs) on lattices provides perhaps
the simplest geometrical model of equilibrium critical phenomena ( i.e.,
non-trivial power-laws in the behavior of different quantities in a system
in thermal equilibrium). The  two other familiar examples of geometrical
models showing phase transitions, the percolation problem and a system of
hard particles (spheres or rods), both involve more complex geometrical
structures. The model of SAWs captures the important macroscopic features
of polymers in solution, and is closely related to other models of
phase-transitions in statistical physics like the Ising model
\cite{ising}, and can also be seen as the $n \rightarrow 0$ limit of the
$n$-vector model \cite{degennes}.  Given the many technological
applications of polymers, and the importance of SAWs as a model of
critical phenomena, it is not surprizing that the model has attracted a
lot of attention in the last sixty years.  Several good reviews are
available summarizing our current understanding of this problem
\cite{reviews}.

The SAW problem is clearly trivial in one dimension. In spite of the large
number of papers related to this problem, an exact solution of the problem
has not been possible so far, for any non-trivial case.  In two
dimensions, the exact value of the growth constant is known for the
hexagonal lattice \cite{hexagon}, and has been conjectured for the square
lattice from large-order exact series expansions \cite{square}. The exact
value of the critical exponents in two dimensions is known from conformal
field theory \cite{duplantier}.  Even for dimensions $d >4$, where the
critical exponents are known to take mean-field values \cite{hara}, a full
solution has not been possible.

Given the paucity of exact solutions in this area, it seems reasonable to
look for some artificially constructed graphs, e.g. fractals, for which
an exact solution can be found. This solution then, can be considered as 
an
approximation to the original problem.  The advantage of this approach,
over other ad-hoc approximations like the Flory approximation, is that one
is assured of well-behavior requirements like the convexity of the free
energy, and avoids problems like getting two different values for a
quantity ( e.g.  the pressure for hard-sphere systems), if one calculates
it in two different ways within the same approximation.

Another motivation for the study of SAWs on fractals comes
from the need to understand how the critical exponents depend on the
dimension and the topological structure of underlying space. In the formal
techniques like $\epsilon$-expansions, one treats the dimension of space
$d$ as a parameter that can be varied continuously, but such techniques involve a formal
analytical continuation of various perturbative field theory integrals of
the type $\int d^d k ..$ to all real values of $d$ \cite{epsilon}.  
However, this formal approach does not give any simple answer to the
question, ``What is the Ising model in 3.99 dimensions ?". On the other
hand, fractals are explicitly constructed spaces of non-integer dimension,
and one can construct fractals with dimension close to any given real
number $d$. One can thus study how critical exponents change by changing
the geometry of the underlying space. Interestingly, the results from the
formal $\epsilon$-expansion techniques do not match with those from
explicitly constructed fractals. The main reason seems to be the
assumption of translational invariance in the former, while the explicitly
constructed fractals do not have this property.  In fact, this assumption
seems to be problematic for spaces with non-integer dimensions, and leads
to pathologies such as non-positive measure \cite{stillinger}. For a
longer discussion of these issues, see \cite{constructive}.

A third reason for the study of SAWs on fractals is that these provide
excellent pedagogical examples of application of renormalization group
techniques to the determination of critical exponents. If one tries to
implement the renormalization transformation to some system like the Ising
model in two dimensions, one eventually generates an infinity of
additional multi-spin or longer range couplings, which are presumably
irrelevant, and are neglected to get renormalization equations in terms of
a finite number of variables. The justification given for doing this
usually involves too much `handwaving' for a discerning student. In case
of fractals, one gets exact renormalization equations in terms of a finite
number of coupling constants. One can then study these in detail, and
explicitly work out their trivial and non-trivial fixed points, stable and
unstable directions, basins of attractions of fixed points, critical
exponents in terms of eigenvalues of the linearized transformation etc.,
and learn the use of renormalization group techniques for determining
critical properties.

Finally, one can treat the fractal lattice as a simple model for a
disordered substrate on which the polymer is adsorbed, and use the exact
results found for polymers on fractals to develop some understanding about
real experimental systems. But for this, this article would not have been
included in this volume.

\begin{figure} 
\begin{center} 
\includegraphics[width=10cm,angle=0]{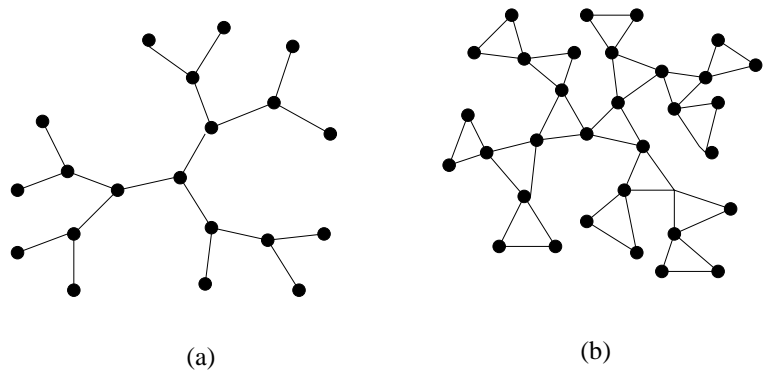} 
\caption{ (a) The Bethe lattice of coordination number $3$. (b) A Husimi 
cactus made up of triangles. } 
\label{husimi}
\end{center} 
\end{figure}

 The simplest of graphs for which the SAW problem is analytically
tractable is the Bethe lattice, for which the exact solution is trivial,
as the graph has no loops.  Only slightly more complicated is the case of
Husimi cactus graphs, which are like the Bethe lattice on the large scale,
but have small loops [Fig. \ref{husimi}]. These do change some properties
like the connectivity constant for the walks, but the large-scale
properties of walks on such lattices are the same as on the Bethe lattice.
In this chapter, we will discuss fractal graphs which have loops of
arbitrarily large size, and for which the critical exponents are different
from the mean-field values.

 This chapter is organized as follows: In section 2, we start with the
precise definitions of the fractals we study. We also define the SAW
generating functions, their annealed and quenched averages and the
critical exponents of self-avoiding walks.  In section 3, we discuss the
renormalization group treatment of the SAWs on the 3-simplex lattice in
detail. We explicitly construct the recursion equations, and show how the
critical exponents of SAWs can be determined from an analysis of the
linearized renormalization group transformation near the fixed point. The
dense phase of the polymer, and the fluctuations of the number of rooted
walks are also discussed briefly.  In section 4, we show how this
treatment is generalized to other fractals where one has more complicated
recursion equations than the simplest case of $3$-simplex lattice..  We
also discuss the behavior of critical exponents on the $n$-simplex in the
limit of large $n$.  In section 5, we discuss the SAWs on the
Given-Mandelbrot family of fractals, and study the behavior of critical
exponents of walks when the fractal dimension of the lattice tends to 2
from below. We use finite-size scaling theory to determine how the
structure of the renormalization equations depends on the parameter $b$
defining the fractals. We develop a perturbative expansion for the
critical exponents valid for large $b$ when the fractal dimension of the
lattice is just below $2$.  In section 6, we discuss the collapse
transition of linear polymers with attractive self-interaction, and the
tricritical $\theta$-point. We also discuss other intermediate
`semi-compact' phases that is seen on some fractals. In section 7, we show
that much of our treatment of linear polymers can be extended to branched
polymers. One can determine the critical exponents $\theta$ for average
number of branched polymer configurations per site, and  $\nu$ 
for the  mean polymer
size  using the RSRG techniques. Interestingly, one finds that for
$GM_b$ fractals with $b>2$, the average number of branched polymers per
site increases as $\exp( a n + b n^{\psi})$, with $0 < \psi < 1$, and the
leading form of correction to the exponential growth is not a power-law
correction. This stretched exponential form arises from the presence of
favorable and unfavorable regions on the fractal lattice.  In section 8,
we discuss a polymer with self-attraction near an attractive surface. In
this case, there is a competition between the collapse transition in the
bulk of the fractal, and the tendency of the polymer to stick to the
surface. We discuss the qualitative features of the phase-diagram, and
critical exponents. The phase-diagram is somewhat complicated when the
next layer interaction is included, and provides a good pedagogical
example for the study of higher order multi-critical points.  One can
study even more complicated systems.  Section 9 contains a discussion of
two mutually interacting linear chains, as a model of the zipping
-unzipping transition in the double-stranded DNA molecules.  Section 10
contains some concluding remarks, and we mention some open problems which
deserve further study.

\section{DEFINITIONS}

We start by giving definitions of the family of fractals that we  
study in this article. All these fractals have a finite ramification number, 
and are defined recursively.

\begin{figure}
\begin{center}
\includegraphics[width=14cm,angle=0]{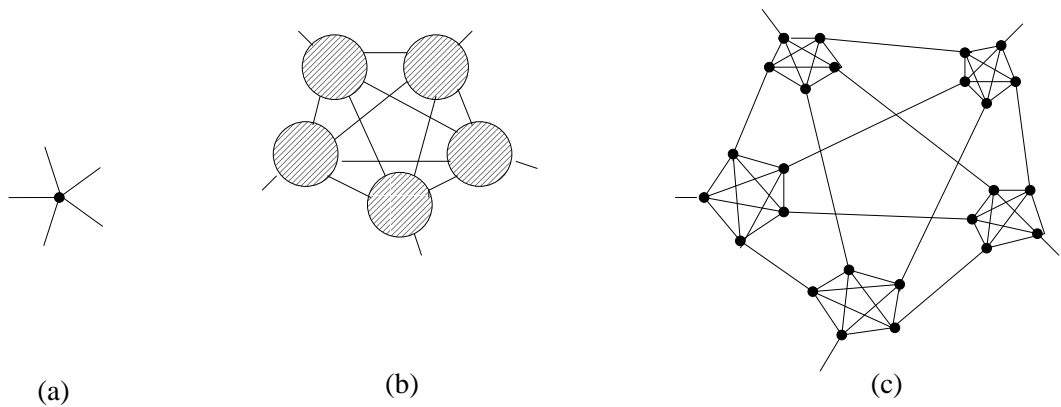}
\caption{ The recursive construction of the $n$-simplex fractal for
$n = 5$. (a) The first order graph (b) the graph of a $(r +1)$
order triangle, formed by joining $n$ $r$-th order graphs, shown as  
shaded blobs here (c) the second order graph. }
\label{nsimplex}
\end{center}
\end{figure}

The first family we shall discuss is the $n$-simplex family, defined for
all positive integers $n >2$. The $n=3$ case was first defined by Nelson
and Fisher \cite{nelson}, who called it truncated tetrahedron lattice.  
The construction was generalized for arbitrary $n$ in \cite{dd1}. The
recursive construction of the graph of the fractal is shown in Fig.  
\ref{nsimplex}. The first order graph is a single vertex with $n$ bonds.
In general, the $r$-th order graph will have $n^{r-1}$ vertices, and 
$(n^r + n)/2$ bonds.  Of these $(n^r -n)/2$ are internal bonds, and there are $n$
external bonds, that are used to connect this graph to other vertices to
form bigger graphs. To form a graph of $(r+1)$-th order, we take $n$
graphs of $r$-th order, and join them to each other by connecting a
dangling bond of each to a dangling bond of the other $r$-order
subgraphs. There is one dangling bond left in each of the subgraphs, and
$n$ bonds altogether. These form the dangling bonds of the $(r+1)$-th
order graph.  We let $r$
tend to infinity to get an infinite graph.  The fractal dimension $D_n$ 
of this graph is easily seen to be $\log n/\log 2$. It was shown in
\cite{dd1} that the spectral dimension ${\tilde d}_n$ of this graph is $2
\log n/ \log (n+2)$.

\begin{figure}
\begin{center}
\includegraphics[width=12cm,angle=0]{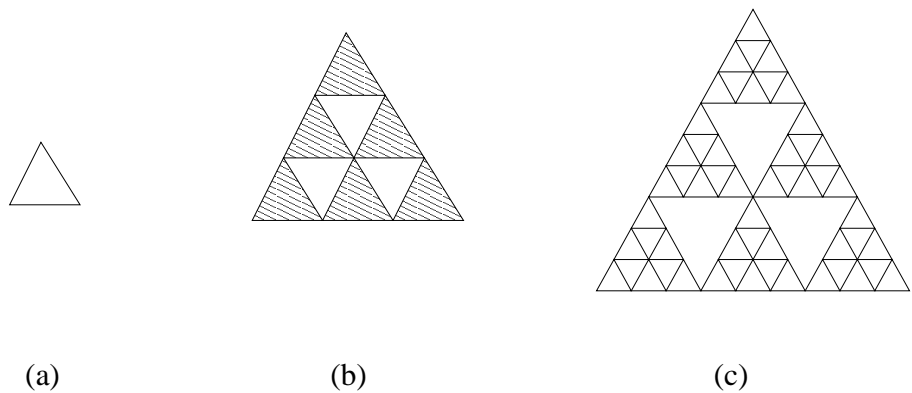}
\caption{ The recursive construction of the Given-Mandelbrot fractal for
$b =3$. (a) The graph of  first order triangle.(b) the graph of a $(r +1)$
order triangle, formed by joining $b(b+1)/2$  $r$-th order triangles
shown as   shaded triangles here (c) graph of the $2$nd order triangle. }
\label{given3}
\end{center}
\end{figure}

The second family we shall consider is the Given-Mandelbrot family of 
fractals \cite{given}, defined in Fig. \ref{given3}. Members of the family are 
characterized by an integer $b$, with  $2 \leq b < \infty$.   
We start with a graph with three vertices and three edges forming a
triangle. This is called the first order triangle. To construct the graph
of the $(r + 1)$-th order triangle, we join together graphs of $b(b +
1)/2$ triangles of $r$-th order, (i.e.  identify corner vertices) as shown
in the figure, to form an equilateral triangle with base which is $b$
times longer. We shall call the resulting graph the $GM_b$ fractal.

It is easy to see that the number of edges in the graph of the $r$-th
order triangle is \\$3 [b (b + 1)/2] ^{r -1}$. The fractal dimension is $
D_b = {\log}[b (b + 1)/2]/\log b$. For $b = 2,3, 4..$, these values are
$1.5849, 1.6309, 1.6609...$ respectively.  The spectral dimension
$\tilde{d}_b$ of the graph can also be calculated exactly for general $b$
\cite{ddspectral}. The values of ${\tilde{d}_b}$ for $b = 2$ to $10$ are
listed in \cite{hilfer}. For large $b$, $\tilde{d}_b$ tends to $2$, and
the leading correction to its limiting value is given by $\tilde{d}_b
\approx 2 - \log \log b/\log b $ \cite{hes}.

\begin{figure}[ht]
\begin{center}
\includegraphics[width=4.0in,height=2.0in]{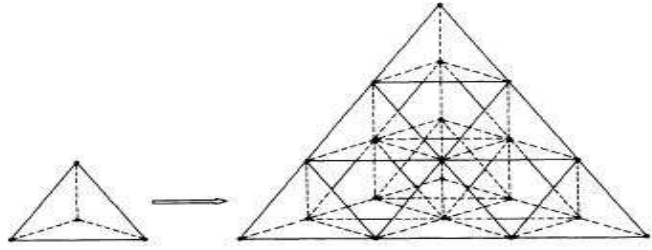}
\caption{First two stages in the iterative construction of the
Hilfer-Blumen fractal $HB(3,3)$.   This  has a fractal
dimension ${\bar d} =\log 10/\log 3.$}
\label{hilferfig}
\end{center}
\end{figure}

In fact, following Hilfer and Blumen \cite{hilfer}, one can define a
general fractal family of Sierpinski -like fractals, to be called
$HB(b,d)$ family here, such that the the $n$-simplex corresponds to $HB(2,
n-1)$, and the $GM_b$ fractal corresponds to $HB(b,2)$.  Here the basic
unit is a $(d+1)$-simplex graph, and one makes the $(r+1)$-th order graph
by making a simplex of $b$ layers of the $r$-th order graphs.  The
construction is illustrated in Fig. \ref{hilferfig} for the fractal
$HB(3,3)$.

\begin{figure}
\begin{center}
\includegraphics[width=10cm,angle=0]{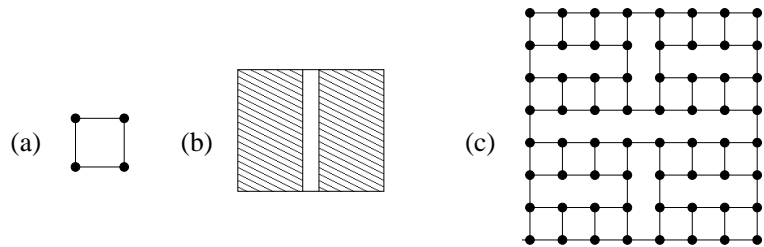}
\caption{ The recursive construction of the  modified rectangular lattice 
for $d =2$. (a) The graph of  first order square.(b) the $(r +1)$-th
order graph, formed by joining $2$  $r$-th order graphs shown as   shaded 
rectangles here (c) graph of the $5$th order rectangle. }
\label{mrl}
\end{center}
\end{figure}

The third family of fractals we shall use is a generalization of the
modified rectangular lattice \cite{dd1}. Here the first order graph is a
set of $2^d$ vertices forming a $d$-dimensional hypercube. In general, the
$r$-th order graph has $2^{r+ d -1}$ vertices in the shape of a cuboid,
with each corner vertex of the cuboid having an extra dangling bond to
connect to outside. The graph of an $(r+1)$-th order cuboid is formed by
taking two $r$-th order cuboids, and joining the opposite faces by
$2^{d-1}$ bonds. The direction of the face that is selected for this is
changed cyclically at the next order, so that the lengths of different
sides of the cuboid are within a factor 2 of each other. It is easily seen
that the fractal dimension of this graph is $d$, and it can be shown that
the spectral dimension is $2( 1 - 2^{ -d})$. The construction is
illustrated in Fig. \ref{mrl} for the case $d = 2$.

The determination of the generating function for the linear polymers on
these fractals follows the treatment of \cite{dd2,rammal}.  Let
$P_n(N)$ be the number of distinct simple polygons  of  perimeter $n$ 
on a graph with $N$ total number of sites in the graph, different
translations of the polymer being counted as distinct. [The use of same 
symbol $n$ for the $n$-simplex and $n$-stepped walk should not cause any 
confusion, as it  is clear from context which is intended.]  
For large $N$, this number increases linearly with $N$. We then define
${\bar P}_n$ as the average number of polygons of perimeter $n$  per site 
by 
\begin{equation}
\bar{P_n} = Lim_{N \rightarrow \infty} \frac{P_n (N)}{N};
\end{equation}
and
\begin{equation}
P(x) = \sum_{n=3}^{\infty} {\bar P}_n x^n,
\end{equation}
For large $n$, ${\bar P_n}$ varies as $ \mu ^n n^{\alpha -3}$, where 
$\mu $ is a constant, and 
$\alpha$ is a critical exponent for the walks. This corresponds to $P(x) 
\sim (1 - \mu x)^{2 -\alpha}$, for $x$ tending to $1/\mu$ from below.
Note that $P(x)$ is a finite-degree polynomial in $x$ for the graphs like the Husimi 
cactus (Fig. \ref{husimi}).

We similarly define $ {\bar C_n}$, the average number of open walks of 
length $n$, as the average over all positions of the root, of the number of 
$n$-stepped self-avoiding walks. For large $n$, this varies as $\mu ^n 
n^{\gamma -1}$, which defines the critical exponent $\gamma$. The 
corresponding generating function
\begin{equation}
C(x) = \sum_{n=1}^{\infty} {\bar C_n} x^n
\end{equation}
varies as $(1- x \mu)^{-\gamma}$ as $x$ tends to $1/\mu$ from below.
We also define the critical exponent $\nu$ for SAWs by the relation that 
the average end-to-end distance for walks of length $n$ varies as $n^{\nu}$ 
for large $n$.

Here in the context of disordered systems, we imagine that the polymer can 
freely move over all space, and the averages calculated correspond to 
annealed averages, as we are  averaging the partition function of SAWs over 
different positions of the root. The logarithm of the number of polymer 
configurations is the entropy, and one can also define the 
equivalent of quenched averages, where one averages, not the numbers of 
walks, {\it but logarithms of numbers of rooted walks}, over different 
positions of the root. These are more difficult to determine. We will
indicate how these can be handled in sec. 3.

\section{RENORMALIZATION  EQUATIONS FOR THE $3$-SIMPLEX FRACTAL}

The procedure to determine the critical behavior of SAWs on fractals is
simplest to illustrate using the $3$-simplex as an example. We will keep 
the presentation informal (hopefully not inaccurate). Readers who prefer a 
more formal approach, may consult \cite{hattori}. We discuss the 
calculation of 
annealed averages first.

\subsection{Calculation of annealed averages}

\begin{figure}
\begin{center}
\includegraphics[width=10cm,angle=0]{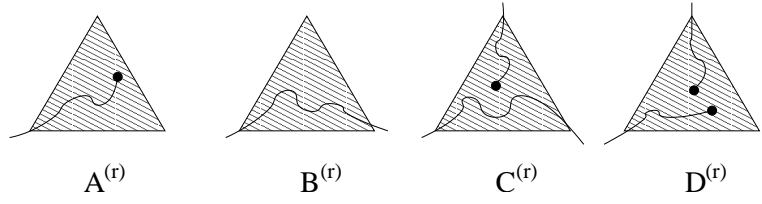}
\caption{ Restricted partition functions for the $3$-simplex lattice. }
\label{3simplex1}
\end{center}
\end{figure}

Consider one $r$-th order triangular subgraph of the infinite order graph. It is
connected to the rest of the lattice by only three bonds. Our aim to sum
over different configurations of the SAW on this subgraph, with a weight
$x$ for each step of the walk. These
can be divided into four classes, as shown in Fig. \ref{3simplex1}. Here 
$A^{(r)}$ is the sum over all configurations of the walk within the 
$r$-th order triangle, that  enters the 
triangle from a specified corner, and  with one endpoint 
inside the triangle. $B^{(r)}$ is the sum over all configurations of walk 
within the triangle that enters and leaves the triangle from specified 
corners. In $C^{(r)}$, 
the walk enters and leaves the subgraph from specified vertices, and 
reenters afterwards from the third corner, and has one endpoint inside the 
triangle. $D^{(r)}$ consists of sum over configurations that have both 
endpoints of the walk inside the triangle, but part of the walk is outside 
the triangle. We can write down the values of these for $r=1$ immediately.

\begin{equation}
A^{(1)} = \sqrt{x}, ~B^{(1)} = x, ~C^{(1)} =~ D^{(1)} =0.
\label{eq4}
\end{equation}

We define the order of a closed polygon on the infinite $3$-simplex graph
to be $r$, if the polygon is completely contained inside an $r$-th order
triangle, but not inside any $(r-1)$-th order triangle. It is easy to see
that sum over all $r$-th order polygons within one such triangle is
${B^{(r-1)}}^3$.  The number of sites in the $(r+1)$-th order triangle is
$3^r$, hence we get

\begin{equation}
P(x) = \sum_{r=1}^{\infty} 3^{-r}  {B^{(r)}}^3.
\label{eq5}
\end{equation}

The sum over open walks can be expressed similarly
\begin{equation}
C(x) =  \sum_{r=1}^{\infty} 3^{-r} [ 3 {A^{(r)}}^2 + 3 B^{(r)} 
{A^{(r)}}^2
+ 3 {B^{(r)}}^2 D^{(r)} ].
\label{eq6}
\end{equation}
   
\begin{figure}
\begin{center}
\includegraphics[width=8cm,angle=0]{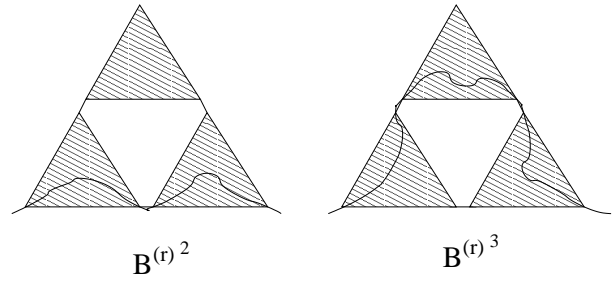}
\caption{ Graphical representation of terms that contribute to the 
recursion equation for $B^{(r+1)}$ for the $3$-simplex lattice. }
\label{3simplex2}
\end{center}
\end{figure}

It is straight forward to write down the recursion equations for these 
weights $A^{(r+1)}, B^{(r+1)},$ $ C^{(r+1)}$ and $D^{(r+1)}$ in terms of 
the 
values at order $r$. For example, Fig. \ref{3simplex2} shows the only two 
possible ways one can construct a polymer configuration of type $B$. It is 
easy to verify that the resulting equations are \cite{dd2}
\begin{equation}
A^{(r+1)} = A ( 1 + 2 B + 2 B^2) + C ( 2 B^2);
\label{eq7}
\end{equation}
\begin{equation}
B^{(r+1)} = B^2 + B^3;
\label{eq8}
\end{equation}
\begin{equation}
C^{(r+1)} = A B^2 + C ( 3 B^2);
\label{eq9}
\end{equation}
\begin{equation}
D^{(r+1)} = ( A^2 +2 A^2 B + 4 A B C + 6 B C^2) + D ( 2 B + 3 B^2);
\label{eq10}
\end{equation}
where we have dropped the superscripts $(r)$ in the right-hand side of the 
equations to simplify notation.

It is straightforward to determine the analytical behavior of the
generating functions $P(x)$ and $C(x)$ from the  equations (4)-(10).

We note that the equation for $B^{(r+1)}$ depends only on $B^{(r)}$, and does
not involve the other variables. The recursion equation (\ref{eq8}) has
two trivial fixed points $B^{*}=0$, and $B^{*}=\infty$, and a non-trivial
fixed point $B^{*}$ given by $B^{*} = (\sqrt{5} - 1)/2$. For $ B^{(0)} = x
< B^{*}$, the variables $B^{(r)}$ decrease with $r$ under iteration, and
tend to zero for large $r$. For $B^{(0)} > B^{*}$, $B^{(r)}$ increases 
with $r$,
and diverges to infinity, and the series for $P(x)$ and $C(x)$ diverge.
This implies that 
\begin{equation}
x_c = B^{*} = 1/\mu = [\sqrt{5} -1]/2.
\end{equation}
which shows that  the growth constant $\mu$
of SAWs on this lattice is the golden mean.

Using Eq. (\ref{eq8}), we see that $P(x)$ satisfies the functional 
equation
\begin{equation}
P(x) = \frac{x^3}{3} + \frac{1}{3} P(x^2 + x^3).
\end{equation}
From this equation, we get $P(x = B^{*}) = \mu^{-3}/2$. For $x$ near 
$B^{*}$, we can linearize the recursion equation. We write $x = B^* - 
\delta $, and $Q( \delta)  = P( B^*) - P( B^* -\delta )$ giving
\begin{equation}
Q( \delta) \approx  \mu^{-2} \delta  + \frac{1}{3}  Q( ( 2 + \mu^{-2}) 
\delta ). 
\label{eq12}
\end{equation}
If we assume that the singular part of $Q(\delta)$ varies as 
$\delta ^{2 -\alpha}$, this gives us
\begin{equation}
\alpha = 2 - \frac{\log 3}{\log (2 + \mu^{-2})} \simeq 0.7342.
\end{equation}

\begin{figure}
\begin{center}
\includegraphics[width=16cm, height=8cm,angle=0]{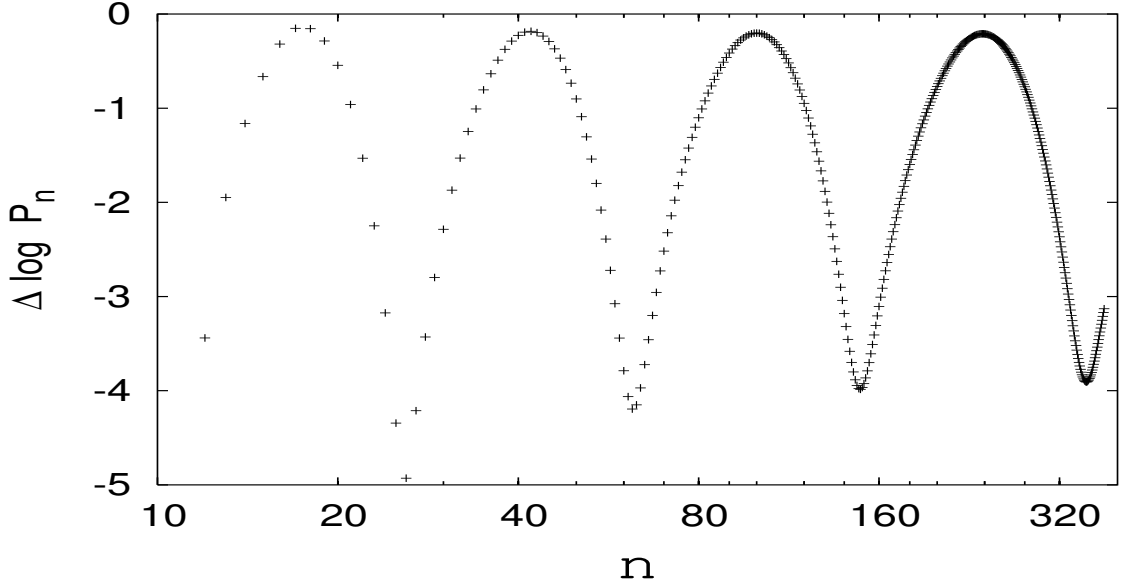}
\caption{ Plot of the difference between the exact value of $\log 
{\bar P}_n$  and the asymptotic form showing log-periodic oscillations 
with $n$ for the $3$-simplex lattice. }
\label{logperiodic}
\end{center}
\end{figure}

The recursion equations for $A^{(r+1)}$ and $C^{(r+1)}$ are linear in 
$A^{(r)}$ and $C^{(r)}$, with coefficients that depend on $B^{(r)}$. If we 
start with a value $\delta \ll 1$, then there is a value $r_0$, such 
that  for $r - r_0 \ll -1 $, we can assume $B^{(r)} \simeq B^{*}$, and for 
$r - r_0
\gg 1$, $B^{(r)} \simeq 0$. As
\begin{equation}
B^{(r+1)} - B^{*} \simeq ( 2 + \mu ^{-2}) ( B^{(r)} - B^{*}),
\end{equation}
we have clearly
\begin{equation}
r_0   = \frac{\log (1/\delta)}{\log ( 2 + \mu^{-2})}.
\end{equation}
For $r < r_0$, in the recursion equations for $A^{(r)}$ and $C^{(r)}$, we 
can put $B^{(r)}= B^*$, giving
\begin{equation}
\left(\begin{array}{ll}  A^{(r+1)}\\  C^{(r+1)}\end{array}\right)
=
\left(\begin{array}{lrlr} ~~1 + 2 \mu^{-1} + 2 \mu^{-2} &  2 \mu^{-2} \\ 
~~\mu^{-2}  &  3 \mu^{-2}
\end{array}\right)
\left(\begin{array}{ll}  A^{(r)}\\  C^{(r)}\end{array}\right).
\label{eq16}
\end{equation}
This implies that $A^{(r)}$ and $C^{(r)}$ vary as $\lambda_{+}^r$, where 
$\lambda_{+}$ is the larger eigenvalue of the $2 \times 2$ matrix in Eq. 
(\ref{eq16}). For $r > r_0$, $A^{(r)} \simeq A^{(r_0)}$, and $C^{(r)} \simeq 
0$, then, in Eq. (\ref{eq6}), the leading contribution comes from the term 
$r = r_0$, giving
\begin{equation}
C(x= x_c -\delta) \simeq (\frac {\lambda_{+}^2}{3})^{r_{_0}} \sim
(1/\delta)^{-\gamma},
\end{equation}
with 
\begin{equation}
\gamma = \frac{ \log (\lambda_{+}^2/3 )} { \log ( 2 + \mu^{-2})} \simeq 
1.3752.
\end{equation}
The typical size of the polymer for $x = x_c - \delta$ is $2^{r_{_0}}$, 
which 
varies as $(1/\delta)^{\nu}$, with
\begin{equation}
\nu = \frac {\log 2}{ \log ( 2+ \mu^{-2})} \simeq 0.7986.
\end{equation}

It is quite straightforward to generate the exact series $P(x)$ using
Eq.(\ref{eq5}) with symbolic manipulation programs like Mathematica.  
Just keeping terms up to $r = 7$, we get a 
series for $P(x)$ exact up to $x^{383}$ ( the coefficient of this term 
has more than 70 digits!). As seen above, the singular
part of $P(x)$ varies as $(1-x^{\mu})^{2-\alpha}$ for $x$ near
$\mu^{-1}$ its $n^{th}$ Taylor coefficient must vary as $\mu^n 
n^{\alpha-3}$.
 However, in general, the solution 
of Eq.(\ref{eq12}) allows an additive term which is periodic function of 
$\log \delta $ with period $\log ( 2 + \mu^{-2})$.
In Fig. \ref{logperiodic}, we have plotted the difference 
$\Delta \log P_n$  between the exact value of $\log {\bar P}_n$ and the 
asymptotic value $ \log (\mu^n n^{\alpha -3})$. We clearly see the 
log-periodic oscillations. Such log-periodic oscillations are seen in many 
other systems showing  discrete scale-invariance \cite{sornette}. The 
existence of these oscillations makes it very difficult to estimate 
critical exponents from extraplations of exact series expansions using 
only a few terms \cite{riera}.

For the fugacity $x > x_c$, the linear polymer fills the 
available space with finite density. Since the logarithm of the single loop 
partition  function for the $r$-th order triangle is now proportional to the number of 
sites in the triangle, we define the free energy per site in the dense 
phase as
\begin{equation}
f(x) = \lim_{r \rightarrow \infty} \frac {\log B^{(r)}}{ 3^{r-1}}.
\end{equation}
Then, from Eq.(\ref{eq8}), it follows that $f(x)$ satisfies the equation
\begin{equation}
f(x) = \frac{1}{3} f(x^2 + x^3).
\label{eq22}
\end{equation}
The density of the polymer $\rho(x)$ in the dense phase is given by
\begin{equation}
\rho(x) = x\frac{d}{dx} f(x) = \frac{ 2+ 3 x }{3 ( x+1) } \rho( x^2 + 
x^3).
\end{equation}
Iterating this equation, we get
\begin{equation}
\rho(x) = \prod_{r = 1}^{\infty} [(  B^{(r)} + \frac{2}{3})(  B^{(r)}+ 
1)^{-1}].
\end{equation}

From Eq. (\ref{eq22}), it can be seen that as $x$ tends to $x_c$ from 
above, the polymer density decreases as $(x - x_c)^{1-\alpha}$. Equivalently, 
for small densities $\rho$, the entropy per monomer 
$\mu(\rho) \simeq  \mu(0) - K \rho^{\frac{1}{1 - \alpha}}$, where $K$ is some constant.

\subsection{Quenched averages}

We note that the $3$-simplex lattice is not homogenous, and different
sites are not equivalent.  In the context of disordered systems, the
calculation of $P(x)$ and $C(x)$ are examples of annealed averages. We 
now show how one can calculate ``quenched averages" in this problem. 

We define $P_n(S)$ as the number of polygons of perimeter $n$ that pass
through a given site $S$ ( these are called rooted polygons, one could
also study rooted open walks), and ask how does it vary with $S$. What is
its average value, variance etcetra ? A quantity of particular interest is
the average value of $\log P_n(S)$.  A good understanding of these for the
regular fractals is prerequisite for understanding the more complicated
case of random quenched disorder. We mention briefly some very recent
results about the behavior of fluctuations of $P_n(S)$ with $S$
\cite{sumedha}.

\begin{figure}
\begin{center}
\includegraphics[width=15cm,angle=0]{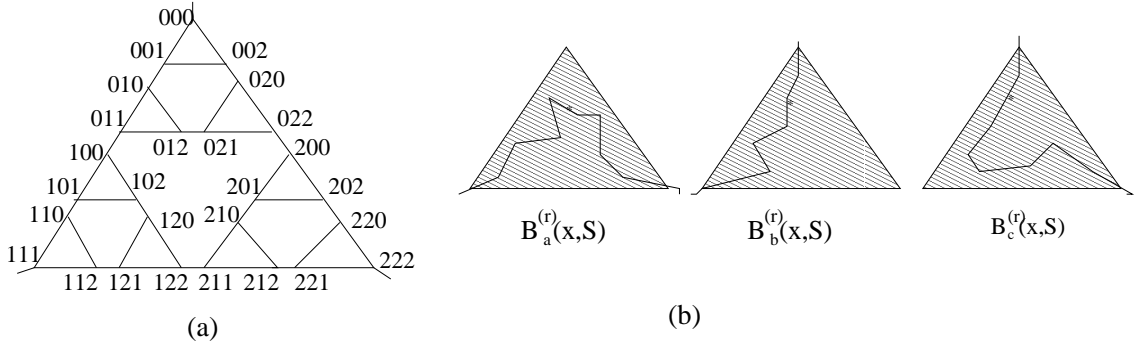}
\caption{(a) labelling of different sites on the $3$-simplex (b) 
definition of weights for rooted graphs }
\label{quenched1}
\end{center}
\end{figure}

To describe rooted polygons, we first need to set up a coordinate system
to describe different sites on the fractal. A simple and natural choice
for this is shown in fig. \ref{quenched1}. A point on the $r$-th order
triangle is labelled by a string of $(r-1)$ characters, e.g.  $0122201
\ldots$. Each character takes one of three values $0,1$ or $2$. The
leftmost character specifies in which of the three sub-triangles the point
lies ($0$, $1$ and $2$ for the top, left and right subtriangle
respectively).  The next character specifies placement in the
$(r-1)$-order sub-triangle, and so on. The restricted partition functions
for the rooted polygons are also defined in \ref{quenched1}. Here
$B^{(r)}_a(x,S)$ is the sum over walks on the $r$-th order triangle that
go through the left and right corners of the triangle, and also visit the
site characterized by string $S$ inside the triangle.  Other weights are
defined similarly.

\begin{figure} 
\begin{center} 
\includegraphics[width=15cm,height=7cm,angle=0]{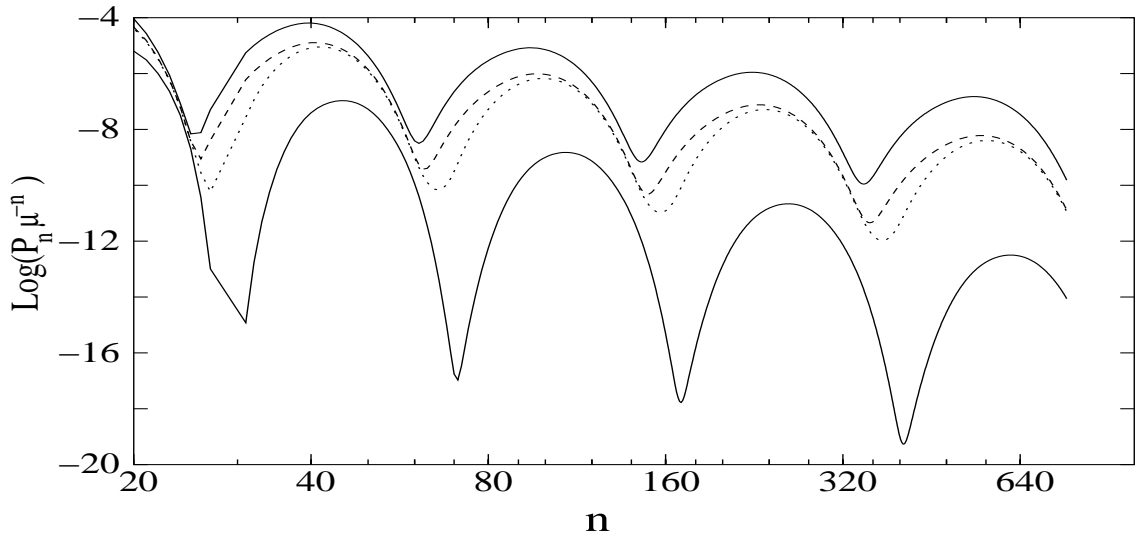} 
\caption{A  plot of the average value of $\log P_n(S) \mu^{-n}$   as a 
function of $n$. The uppermost and the 
lowermost curves are  theoretically derived  upper and lower bounds 
to this number over different positions of the root. The dashed and dotted 
lines show the annealed and quenched average values respectively. } 
\label{quenched2}
\end{center} 
\end{figure}

A site in the $(r+1)$-th order will be characterized by a string of $r$ 
characters.  Hence, a site characterized by string $S$ at the $r$-th stage 
will be characterized by one of the strings $0S$, $1S$ or $2S$. The 
recursion relations in this case are linear, and can be written in the 
matrix form

\begin{equation}
\left(\begin{array}{ll}  B^{(r+1)}_a(x,sS)\\
 B^{(r+1)}_b(x,sS)\\  B^{(r+1)}_c(x,sS)\\ \end{array}\right)
=  {\cal{M}}_s \left(\begin{array}{ll}  B^{(r)}_a(x,S)\\
 B^{(r)}_b(x,S)\\ B^{(r)}_c(x,S)\\ \end{array}\right),
\label{eq24.1}
\end{equation} 
where 
\begin{equation} 
{\cal{M}}_0 =
\left(\begin{array}{lll} B^2 ~~0 ~~0 \\ ~0 ~~B ~~B^2\\ ~0 ~~B^2 ~B\\
\end{array}\right);~~{\cal{M}}_1 = \left(\begin{array}{lll} ~B ~~0 ~B^2 \\
~~0 ~B^2 ~0\\ ~B^2 ~0 ~~B\\ \end{array}\right);~~{\cal{M}}_2 =
\left(\begin{array}{lll} B ~B^2 ~0 \\ B^2 ~B ~0 \\ ~0 ~~0 ~B^2\\
\end{array}\right), 
\label{eq24.2} 
\end{equation} 
and we have suppressed
the superscript $(r)$ on $B$ in the matrices.  The generating function for
rooted polygons is given by 
\begin{equation} 
P(x,S) = \sum_{r=1}^{\infty}
B^{(r)}_a (x, S_r) {B^{(r)}}^2, 
\end{equation} 
where $S_r$ is the string
consisting of the last $(r-1)$ characters of the position of $S$.  If we
ignore the constraint that the walk has to pass through $S$, we get an
upper bound on the number of such walks, $B^{(r)}_a(x, S) \le B^{(r)}(x)$,
where inequality between polynomials is understood to imply inequality for
the coefficient of each power of $x$.  This implies that for all sites
$S$, \begin{equation} P(x,S) \leq \sum_{r=0}^{\infty} {B^{(r)}}^3.
\end{equation} If we write the right-hand-side as ${\cal F}_u(x)$, then
${\cal F}_u(x)$ satisfies the equation \begin{equation} {\cal F}_u(x) =
x^3 + {\cal F}_u(x^2 + x^3). \end{equation} This equation should be
compared with Eq.(\ref{eq12}), and differs from it only in the absence of
the factor $1/3$. This functional equation has the fixed point at $x^* =
1/\mu$, and linear analysis near the fixed point shows that ${\cal
F}_u(x)$ diverges as $- \log (1-x \mu)$ for $x$ tending to $1/\mu$ from
below. This implies that the coefficient of $x^n$ in the Taylor expansion
of ${\cal F}_u$ varies as $\mu^n/n$ for large $n$. Thus, \begin{equation}
P_n(S) \leq K \mu^n/n, {\rm ~~for ~all } ~n, {\rm ~and ~all } ~S,
\end{equation} where $K$ is some constant.  Sumedha and Dhar
\cite{sumedha} also derive a lower bound $P_n(S) \geq K_1 \mu^n n^{-b-1}$,
for all $n$ and all $S$, where $b = 2 \log \mu/ \log ( 2 + \mu^{-2})
\simeq 0.92717$, and $K_1$ is some constant. For a randomly chosen $S$,
the problem reduces to a random product of noncommuting matrices ${\cal
M}_0$, ${\cal M}_1$ and ${\cal M}_2$.  The expected value of the logarithm
of a product of $r$ independent random matrices varies linearly with $r$
for large $r$.  This implies that for large $n$, the quenched average
$<\log P_n(S)>$ varies as $ n \log \mu - (2 -\alpha_q) \log n$. The
numerical estimate of $\alpha_q$ is approximately $0.729 \pm .001$, which
is just a bit less than the annealed value $0.7342$. In figure
\ref{quenched2}, we have plotted the numerical values of the upper and
lower bounds to $\log (P_n(S) \mu^{-n})$, and  the annealed and 
quenched averages 
$\log < P_n(S) \mu^{-n}>$, and $< \log (P_n(S) \mu^{-n})>$.  It should be
noted that the lower bound is the best possible, in the sense that for
each value of $n$, there is a finite density of roots that saturate the
bound.  The upper bound is not always optimal.  Also, whether the annealed
and quenched averages are very close or less so appears to be an 
oscillatory function of $\log n$.

\section{RECURSION EQUATIONS FOR OTHER FRACTALS}
    
We now briefly indicate how the real-space renormalization group 
technique of the previous section can be extended to other fractals.

\subsection{$n$-simplex fractals for $n>3$}

The case of $n$-simplex fractal for higher $n$ is a straight forward 
extension of the technique used for the $3$-simplex. The case $n= 4$  
was studied   in \cite{dd2}, and the treatment was extended to 
$n=5$ and $6$ in  \cite{ksj}.

\begin{figure}
\begin{center}
\includegraphics[width=15cm,angle=0]{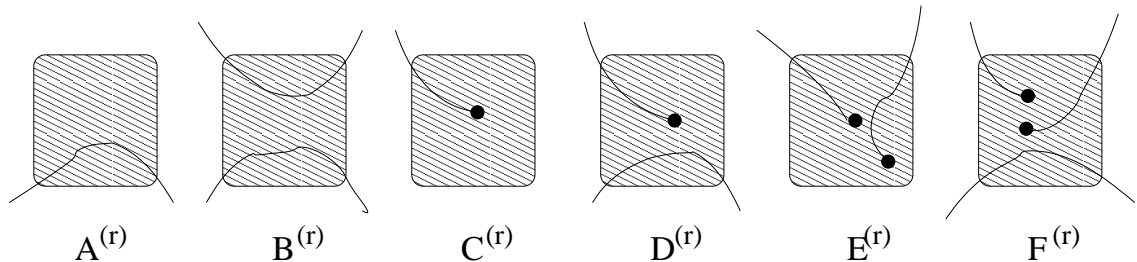}
\caption{Different restricted partition functions  for the $4$-simplex 
lattice. The shaded squares denote graphs of $r$-th order $4$-simplex, of 
which only the external bonds are shown. }
\label{4simplex}
\end{center}
\end{figure}

For higher $n$, the number of restricted partition functions we have to 
define to generate a closed set of recursion equations is larger. Here 
the permutation symmetry between the corner vertices of the $n$-simplex 
graph is very useful in reducing the number of variables needed.  For 
$n=4$ and $5$, we need two functions $A^{(r)}$ and $B^{(r)}$ to generate 
the closed loops generating function $P(x)$. For the open walks 
generating function $C(x)$, four more variables are needed.   Their 
definition is shown in Fig. \ref{4simplex}.
They will be denoted by $C^{(r)}, D^{(r)}, E^{(r)}$ and $F^{(r)}$.

The starting values of these weights are 
\begin{equation}
A^{(1)}=x;\hspace{2cm} C^{(1)}=\sqrt{x};\hspace {2cm} 
B^{(1)}=D^{(1)}=E^{(1)}=F^{(1)}=0.
\end{equation}
The recursion relations for these weights are written by
constructing graphs by all possible ways. This leads to \cite{dd2}
\begin{equation}
A^{(r+1)} = A^2+2A^3+2A^4+4A^3B+6A^2B^2, 
\label{eq26}
\end{equation}
\begin{equation}
B^{(r+1)} = A^4+4A^3B+22B^4, 
\label{eq27}
\end{equation}
and similar equations for the other variables (omitted here).
The expression  for $P(x)$ in this case is  
\begin{equation} 
P(x)=\sum_{r=1}^\infty (4{A^{(r)}}^3+3{A^{(r)}}^4).
\end{equation}
There is a similar but more complicated expression for the generating
function for open walks. We note that recursions of $A^{(r+1)}$ and
$B^{(r+1)}$ depend only on $A^{(r)}$ and $B^{(r)}$. For any given value of
$x$, we can determine $P(x)$ numerically by explicitly iterating the
recursion equations. One finds

\begin{equation} 	 
x_c=0.4373, \qquad   \mu=2.2866,
\end{equation} 	 
and the corresponding non-trivial fixed point is $(A^*,B^*) 
= (0.4264,0.04998)$. Linearization of Eqs. (\ref{eq26}) and (\ref{eq27}) around this fixed 
point gives largest eigenvalue $\lambda_1=2.7965$. We can express the 
end-to-end 
distance exponent $\nu$ in terms of $\lambda_1$ using arguments similar to the
case of 3-simplex. We get
\begin{equation} 	 
\nu=\frac{\log 2}{\log\lambda_1}=0.6740,
\end{equation} 	 
and specific heat exponent
\begin{equation} 	 
\alpha=2- \nu D_4 = 0.6519.
\end{equation}

To find $\gamma$, we consider the recursion equations for $C^{(r+1)}$ and 
$D^{(r+1)}$, which are linear in
$C^{(r)}$ and $D^{(r)}$, and diagonalizing the $2 \times 2$ matrix 
gives the eigenvalue $\lambda_+=4.2069$, which corresponds to
$\gamma=\log[(\lambda_+^2)/4]/\log[\lambda_1] = 1.4461$. 	 

For the $5$-simplex lattice, the closed loop generating function can be 
found in terms of variables $A^{(r)}$ and $B^{(r)}$, defined as before. In 
this case, the recursion equations are more complicated \cite{ksj}. We 
list these here to show how the complexity of the polynomials rises 
rapidly with increasing $n$:
\begin{equation} 	 
A^{(r+1)} = 132 (B^5 +  A B^4) +  A^2 ( 1 + 18 B^2 + 96 B^3) + A^3 ( 3+ 
12 B +78 B^2)+ A^4 ( 6 + 30 B) + 6 A^5,  
\label{eq32}
\end{equation}
\begin{equation}
B^{(r+1)} = 186 B^5 + 220 A B^4 + 88 A^2 B^3 + A^4 ( 1 + 13 B +220 B^4) + 
2 A^5. 
\label{eq33}
\end{equation}

We omit the other equations. Starting with $(A^{(1)},B^{(1)})=(x,0)$, one 
finds a
non-trivial fixed point for $x =x_c = 0.336017$, which corresponds to $
\mu = 1/x_c = 2.97603$ and the fixed point $(A^*,B^*)=(0.3265,0.02791)$.  
Linearization of the recursion relations Eq. (\ref{eq32}) and (\ref{eq33})
around this fixed point leads to one eigenvalue $\lambda_1=3.13199$
greater than unity. Using this value of $\lambda_1$ one gets $\nu=0.6049$
and $\alpha=0.5955$.

The exponent $\gamma$ is found from the recursion relations for the 
variables corresponding to a single end point.  The largest
eigenvalue $\lambda_+$ of the matrix which characterizes the linear
transformation of the recursion relations of these functions is
$\lambda_+=5.24398$ corresponding to $\gamma = 1.4875$.

For the $6$-simplex lattice, even for the polygon generating function, we 
need three restricted partition functions $A^{(r)}, B^{(r)}$ and 
$C^{(r)}$, corresponding the cases where the walk enters and exits the 
$r$-th order subgraph once, twice or three times respectively. For the 
open walks, we need six more variables. For details, consult \cite{ksj}. 
In this case, the nontivial fixed point is found to be  
\begin{equation} 	 
(A^*,B^*,C^*)=(0.262352,0.017588,0.0007011).
\end{equation} 	 

This fixed point is reached if we start with $x = x_c=0.27166$, which
corresponds to the connectivity constant $\mu=3.68107$. The linearization
of recursion relations about this fixed point yields only one
eigenvalue, $\lambda_1=3.4965$, higher than unity giving $\nu=0.5537$ and
$\alpha=0.5686$.

By diagonalizing the matrix corresponding to the linear recursions for
weights corresponding to one end point, one gets the largest eigenvalue
$\lambda_+=6.26709$. This gives  $\gamma=1.500094$ for the  $6$-simplex lattice.

\subsection{$n$-simplex lattice in the limit of large $n$}

One can explicitly determine the critical exponents for a few more values
of $n$ using a computer to generate the recursion equations. However, this
soon becomes  very time-consuming. If $n$ is very large, some simplifications
occur, and one can determine the leading behavior of the critical exponents
of the SAW problem on the $n$-simplex in this limit \cite{Singh1}. We 
discuss this below.

We start by noting that as $n$ increases, the probability of occurrence of
loops in a random walk on the graph decreases, and as the loops become
rarer, the random walk without self-exclusion should approximate the
properties of the SAW. In particular, we would expect that the growth
constant for SAWs on the $n$-simplex should be approximately equal to
$n-1$.  Let us denote, as before, the restricted partition functions for
the $r$-th order lattice corresponding to polygon generating function by
$A^{(r)}, B^{(r)}, C^{(r)} \dots$, corresponding to configurations where
the SAW enters (and exits) the graph once, twice, thrice
$\ldots$respectively. Then for the fixed point corresponding to the
swollen state, we would have $ A^* \sim {1}/{n}$. Since $B^{(r)} \leq 
{A^{(r)}}^2$, it  is at most of order ${1}/{n^2}$, and  similarly, 
$C^* $ is at most of order ${1}/{n^3}$. Thus, $B^*,C^*,.....$ approach 
zero faster than $A^*$ as $n$ is increased.

It is straight forward to write down the recursion equation for 
$A^{(r+1)}$ if we neglect the terms involving $B, C, $ etc. There is only 
one configuration of walk corresponding to the term $A^2$ ( we drop the 
superscript of $A^{(r)}$ for convenience), but $(n-2)$ terms of type 
$A^3$, 
and $(n-2)(n-3)$ terms of type $A^4$ etc. This gives us the recursion 
equation 
\begin{equation} 	 
A^{(r+1)} \simeq A^2+(n-2)A^3+(n-2)(n-3)A^4+.......+(n-2)!A^n.
\label{eq35}
\end{equation}

When $A$ is of order $1/n$, each of these terms is of order $1/n^2$, and 
the sum needs to be evaluated with some care. In \cite{ksj}, it was 
noted that,  with only a small error when $A$ is near $A^*$, this series 
can be rewritten as
\begin{eqnarray}
A^{(r+1)}& \simeq& A^n (n-2)! [ 1 + 1. A^{-1} + (1/2!) A^{-2} +
\dots +( 1/( n-2)!) A^{-n+2}  ] \\
 &\simeq &   (n-2)! ~A^n~ \exp(1/A)
\end{eqnarray} 

One can then study the asymptotic behavior of this recursion equation for 
large $n$. Here we 
shall use a different approach. If we 
change the variable from $A^{(r)}$ to $\epsilon^{(r)}$ using the relation
\begin{equation}
A^{(r)} = \frac{1}{n} \exp ( \frac {\epsilon^{(r)}}{\sqrt{n}}).
\end{equation}
and use the fact that, for $n \gg r \gg 1$, 
\begin{equation}
\prod_{j=1}^{r} ( 1-j/n) \simeq  \exp( -\frac{r^2}{2n}),
\end{equation}
the series of Eq.(35) can be written as
\begin{equation}
A^{(r+1)} \simeq (\frac{1}{n^2}) \sum_{r=2}^{n} \exp ( 
\frac{\epsilon^{(r)}
r}{\sqrt{n}} - \frac{r^2}{2 n}).
\end{equation}

Substituting $x$ for $r/\sqrt{n}$ and replacing the summation over $r$ for 
$\sqrt{n} \gg \epsilon^{(r)} \gg 1$ 
by integration over $x$ from $-\infty$ to $\infty$ (for $\epsilon \gg 1$, the lower limit can be
extended to $-\infty$ with negligible error) we get 
\begin{equation}
A^{(r+1)} \simeq (\frac{\sqrt{ 2 \pi}}{n^{3/2}})  \exp( 
\frac{{\epsilon^{(r)}}^2}{2}).
\end{equation}
From this equation, we see that  the nontrivial fixed point is given by
\begin{equation}
\epsilon^{(r)}= \epsilon^* \approx \sqrt{\log n}.
\label{eq39}
\end{equation}
Let us now look at the equation for $B^{(r+1)}$. Again keeping only the 
terms involving $A$'s alone, we get
\begin{equation}
B^{(r+1)} \simeq A^4 + 2 (n-4) A^5 + 3 (n-4)(n-5) A^6 +\ldots
\end{equation}
Using the same approximation as before, we get
\begin{equation}
B^{(r+1)} \simeq (\frac{1}{n^4})\sum_{r=1}^{n} r  \exp ( \frac{\epsilon 
r}{\sqrt{n}} - \frac{r^2}{2 n}) \simeq \frac{\sqrt{2 \pi \log 
n}}{n^{5/2}},
\end{equation}  
where we have used the approximate fixed point value of $\epsilon^*$ from 
Eq.(\ref{eq39}). The fact that $B^{(r+1)}$ decreases faster than $n^{-2}$ 
justifies neglecting these terms in determining the asymptotic behavior of 
the recursions.
The value of the derivative of the linearized recursion equation for $A$ 
at the nontrivial fixed point is
\begin{equation}
\frac{d}{dA} A^{(r+1)} \simeq \sqrt{ n \log n}.
\end{equation}
This implies that the critical exponent $\nu$ is given by
\begin{equation} 
\nu \simeq \frac {2 \log 2}{ \log n} \left[1-\frac{\log \log n}{\log n} 
+{\rm 
higher ~~order~~ terms} \right].
\end{equation}
 	 
We note that the correction term to Eq.(\ref{eq35}) involving $B^{(r)}$ in
leading order are of the form
\begin{equation}
\Delta A^{(r+1)} = 2A^3B(n-2)(n-3) + 4A^4B(n-2)(n-3)(n-4) + ....
\end{equation}
Using the estimate Eq.(50) for $B^{(r)}$, we see that near the fixed
point, the error in Eq.(41) $\Delta A^{(r+1)}$ is of order $n^{-3/2}$.
This implies that the fractional error in the value of $\nu$ using Eq.(41)  
is of order $n^{-1/2}$. Thus Eq.(41)  is much more accurate than Eq.(52),
where the error is of order $1/(\log n)$.

To  calculate $\gamma$, one has to consider  configurations with one 
endpoint of walk inside the graph. Again, keeping only terms involving 
powers of $A$, the recursion relation for the weight of configurations with the walk 
entering the graph once  ( call it $X$) is found to be 
\begin{equation}
X^{(r+1)}\simeq X [1+(n-1)A^* + (n-1)(n-2)A^{*2} + .... + 
(n-1)!A^{*(n-1)}].
\end{equation}
Using the arguments as before, this can be evaluated using the fixed point 
value of $A^*$, giving
\begin{equation}
X^{(r+1)} \simeq  X  \sqrt{n \log n}.
\end{equation} 	 
It follows that, for large $n$,

\begin{equation} 	  
\gamma\simeq 2 \left[ 1-\frac{\log \log n}{\log n} +{\rm ~higher 
~order~terms}\right].
\end{equation} 	 

Note that the critical exponents do not take mean-field values, even
when the fractal dimension of the lattice becomes greater than $4$. This
is clearly because of the special structure of the $n$-simplex lattice,
where, even though the fractal dimension becomes greater than $4$ for
large $n$, the spectral dimension remains below $2$, and probability of
intersection of paths of random walks remains large. Also that the
non-analytical dependence of the type $\frac{\log \log n}{\log n}$ in the
critical exponents on the lattice cannot be obtained from
$\epsilon$-expansion like power-series expansions in deviation of
dimensionality from some reference value.

\subsection{Modified rectangular lattice}

This lattice is interesting as its fractal dimension, and the spectral
dimension are both rational numbers. Also, one can get its graph by
selectively deleting some bonds from the graph of $d$-dimensional
hypercubical lattice. Since the $(r+1)$-th order graph is formed by taking
only two smaller graphs, the recursion equations involve only a small
number of configurations, and are easy to write down. But the number of
variables needed is larger, as the symmetry of the graph is lower than
that of the $HB(b,d)$ family.

\begin{figure}[ht]
\includegraphics[width=6.0in,height=1.5in]{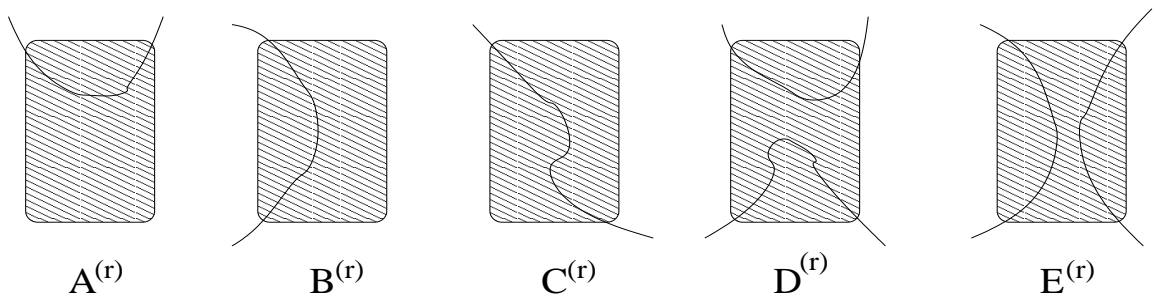}
\caption{Diagrams representing the restricted partition functions for the various ways
the polymer can cross the r$^{th}$ order rectangle.}
\label{mrlfunc}
\end{figure}

 The behavior of SAWs on the $d=2$ lattice was studied in \cite{dd2}.
The recursion equation for a polygon is written in terms of five
restricted generating functions (see Fig. \ref{mrlfunc}) by constructing graphs by all
possible ways \cite{dd2}. This gives

\begin{eqnarray}
A^{(r+1)} & = & B(1+D), \qquad B^{(r+1)} = A^2 + C^2, \qquad C^{(r+1)} = 2AC, 
\nonumber \\
D^{(r+1)} &=& B^2 +2DE, \qquad E^{(r+1)} = D^2.
\end{eqnarray}

The starting values of these recursions are 
\begin{eqnarray}
A^{(1)} &=& B^{(1)} = x^2 + x^4, \qquad C^{(1)} = 2x^3, \qquad D^{(1)} = E^{(1)} = 
x^4.
\end{eqnarray}

The polygon number generating function  $P(x)$  is given by \cite{dd2} 
\begin{equation}
P(x) = \frac{x^4}{4} +  \sum_{r=1}^{\infty} [\frac{(B^{(r)})^2}{2^{r+2}}].  
\end{equation}
Numerical iteration of these equations gives $x_c = 0.5914$ and $\mu = 
1.6909$. The fixed point occurs at 
\begin{equation}
(A^*, B^*, C^*, D^*, E^*) = (0.5000, 0.4201, 0.4124, 0.1902, 0.0362), 
\end{equation}
giving one eigenvalue $\lambda_1 = 1.6839$ greater than one. Using this
eigenvalue one finds $\nu = \log 2/2 \log \lambda_1 = 0.66503 \; {\rm and}
\; \alpha = 0.6699$. A similar analysis for the open-walk configurations
gives the critical exponent $\gamma = 1.4403$.

\section{SELF-AVOIDING WALKS ON FRACTALS WITH DIMENSION $2 - \epsilon$}

In the preceding sections we  studied the properties of SAWs on several
different fractals.  For the calculation of critical exponents of SAWs on
still other fractals, see for example \cite{melrose}. It would appear that
for each fractal, one has to write down the polynomial recursion 
equations, and
calculate the values of critical exponents using the technique outlined.
There is no simple expresion for the critical 
exponents as a function of the geometrical
properties of the fractal, ( an improved Flory formula?) that would allow 
one to predict these {\it without doing the full calculation}.
Unfortunately, this level of understanding of the problem is still not
achieved.

The best way to understand such systematics is to study the variation of
exponents across a family of fractals as some property of the fractal is
changed. For the $n$-simplex family studied earlier, the fractal and
spectral dimensions did not tend to the same value for large $n$. We now
discuss variation of critical exponents for SAWs on the $GM_b$ fractals
family  as the parameter $b$ is varied from $2$ to infinity.
For large $b$, both the fractal and spectral dimensions of the fractal tend
to $2$ where the properties of walks are well-understood. Such a study
makes contact with the formal $\epsilon$-expansion technique which has
played an important role in development of renormalization theory of
critical phenomena.

The self-avoiding walks for the $GM_2$ fractal have the 
same exponents as for the $3$-simplex. For small  values 
of $b$, it is  straight forward to  generate the explicit 
recursion equations on a computer, and determine the critical exponents. 
These were 
worked out  by Elezovic et al \cite{elezovic} for  $b$=$2$ to $8$, and 
these 
studies were extended to $b=9$ by Bubanja et al \cite{bubanja2}. The 
general 
form of recursion equations for the function $A^{(r)}, B^{(r)}, C^{(r)}$
and $D^{(r)}$ ( definition of these is same as in Fig. \ref{3simplex1})
are of the form
\begin{eqnarray}
B^{(r+1)} &=& f_b( B^{(r)}), \nonumber \\ 
\left[\begin{array}{ll}  A^{(r+1)}\\  C^{(r+1)}\end{array}\right]
&=&
\left[\begin{array}{lrlr} ~~P_{11}(B^{(r)}) &  P_{12}(B^{(r)}) \\
~~P_{21}(B^{(r)})  &  P_{22}(B^{(r)})
\end{array}\right]
\left[\begin{array}{ll}  A^{(r)}\\  C^{(r)}\end{array}\right].
\label{eq52}
\end{eqnarray}
where $f_b, P_{11}, P_{12}, P_{21}$ and $P_{22}$ are 
polynomials of $B^{(r)}$, whose exact form depends on $b$.

  From  analysis of these equations, it was found
that the critical exponent $\nu$ for SAWs takes the values $0.7986, 
0.7936, 0.7884$, $0.7840, 0.7803, 0.7772, 0.7744$ and $ 0.77218$ as $b$ is varied 
from $2$ to $9$. It thus seems to converge to the Euclidean value $\nu_{2d} =
3/4$. However, the value of the critical exponent $\gamma$ changes from  
$1.3752, 1.4407, 1.4832, 1.5171, 1.5467$ and $1.5738$ as $b$ is varied
from $2$ to $7$, and seems to diverge away from the known exact two
dimensional value $\gamma_{2d} =43/32 \simeq 1.344$.

The variation of these exponents with $b$ for large $b$ was explained in 
\cite{dd3} using the finite size scaling theory. For large $b$, the growth 
constant of SAWs on the $b$-fractal   would be close to the critical 
value in two dimensions. It is convenient to change variables from 
$B^{(r)}$ to a variable that is proportional to the departure from 
criticality in these systems. We write 
\begin{equation}
B^{(r)} = B_{b =\infty}^{*} \exp( \epsilon^{(r)}).
\end{equation}
Then $B^{(r+1)}$ is related to the properties of SAWs that traverse  an 
equilateral triangle of side $b$ from one corner to another. From 
finite-size scaling theory, this would be expected to be a function of  
a single variable $\epsilon b^{1/\nu}$:
\begin{equation}
f_b(B) \simeq \frac{K}{b^a} \exp[ g( \epsilon b^{1/\nu})],
\label{eq54}
\end{equation}
where $K$ is some constant, and $a$ is an exponent. From the conformal 
field theory \cite{duplantier,cardy}, the scaling dimension of a spin at the 
corner of  a wedge of angle $\pi/3$ is known, and that implies that 
a=15/4. The scaling function $g(x)$ has to have the following 
properties:\\ 
(i) It is a monotonically increasing convex function of $x$. We can 
set $g(0) = 0$, by redefining K. 
(ii) For $\epsilon < 0$, $f_b$ should decrease exponentially with $b$. 
This implies that $g(x) \sim  - |x|^{\nu}$ for $x \ll -1$. 
(iii) For fixed $\epsilon > 0$, $f_b$ should vary as $\exp ( b^2)$ for 
large $b$. Thus, we must have 
\begin{equation}
g(x) \simeq K_1  x^{2 \nu}, {\rm for~} x  \gg1.
\label{eq55}
\end{equation}

From Eqs. (\ref{eq52}), it is easy to see that  the fixed point 
value of $\epsilon$ for large $b$ is given by $g(\epsilon^{*}_b b^{1/\nu}) 
\simeq a 
\log b$. Using Eq.(\ref{eq55}), this gives
\begin{equation}
\epsilon^{*}_b \simeq \left(\frac{a \log b}{K_1}\right)^{1/2\nu} 
b^{-1/\nu}.
\end{equation}
The derivative of $B^{(r+1)}$ with respect to $B^{(r)}$ at the 
fixed point is found to be 
\begin{equation}
\lambda_1(b) = b^{1/\nu} \frac{dg}{dx}|_{ \epsilon = \epsilon^*_b} \simeq 
K_2 b^{1/\nu} (\log b)^{\frac{2 \nu -1}{2 \nu}},
\end{equation}
where $K_2 = 2 \nu a ( K_1/a)^{1/2\nu}$ is a constant. Expressing the 
critical exponent $\nu(b)$ in terms of $\lambda_1(b)$, we get
\begin{equation}
\frac{1}{\nu(b)} = \frac{1}{\nu} + \frac{2 \nu -1}{2 \nu} \frac{ \log \log 
b}{\log b} +\dots,
\end{equation}
where the dots represent terms of order $\frac{1}{\log b}$.

It is interesting to note that the leading correction term to the
asymptotic value of $\nu$ is negative.  Thus this analysis predicts that
variation of $\nu$ with $b$ is not monotonic. In particular, $\nu_b$
should be less than $\nu$ for large enough $b$. This rather striking
prediction of the finite-size scaling analysis given above was
checked by Milosevic and Zivic using numerical Monte-carlo renormalization
group studies \cite{mcrg1,mcrg2}, and they found that this happens
for $b \approx 27$.

We can similarly determine the leading correction to the susceptibility 
exponent $\gamma$.  It is known that in two dimensional critical theory, 
a $3$-leg vertex has a higher scaling dimension than a $1$-leg vertex. 
This implies that at the critical point $\epsilon =0$, the ratio  
$C^{(r)}/ A^{(r)}$ will tend to zero as a  power of $b$, and hence 
$C^{(r)}$  can be 
ignored in determining the leading correction. Then the value of the 
exponent $\gamma$ is determined by the value of $P_{11}$ at the fixed 
point $\epsilon = \epsilon^*(b)$. We can now write the scaling ansatz for 
this variable:
\begin{equation}
P_{11}(\epsilon, b) \simeq K_3 b^c \exp [ h( \epsilon b^{1/\nu})],
\end{equation}
where $K_3$ is a constant, $c$ is a critical exponent, and $h(x)$ is a 
scaling function. 

For $\epsilon < 0$, it is known that as $b \rightarrow \infty$, $P_{11}$ 
varies as $\epsilon^{-1/64}$ \cite{cardy,guttmann}. This implies that we 
must have 
$h(x) \sim (-1/64) \log x$, and $ c= 1/48$.

Now, for large $x$, the functions $h(x)$ and $g(x)$ should increase in a 
similar way. Then using Eq.(\ref{eq54}), we can write $P_{11}$ as
\begin{equation}
P_{11}( \epsilon^*_b,b) \simeq K_4 b^{c+a} \exp[ h(\epsilon^*_b b^{1/\nu}) 
- g(\epsilon^*_b b^{1/\nu})] B^*(b).
\end{equation}
For large $x$, both $h(x)$ and $g(x)$ increase as $K_1 x^{2\nu}$, but 
the leading dependence  is the same exactly. Thus, for large $x$, $h(x) 
-g(x)$ varies at 
most as a sublinear power of $\log b$. Thus, we get
\begin{equation}
\lim_{b \rightarrow \infty} \frac{\log P_{11}(\epsilon^*_b,b)} {\log b} = 
c + a.
\end{equation} 
As $\gamma(b) =  \log[ \frac{2 P_{11}^2}{b(b+1)}]/\log b$,  this 
implies that
\begin{equation}
\lim_{b\rightarrow \infty} \gamma(b) = 2( c+a -1) \nu = 133/32.
\end{equation}
Using the  known scaling exponents for the dense polymer problem in two 
dimensions, it was shown in \cite{dd3} that the leading correction to 
the large $b$ limiting value of $\gamma$ is given by
\begin{equation}
\gamma(b) = \frac{133}{32} - \frac{321}{128} \frac{\log \log b}{\log b} 
+{\rm 
~higher ~order~terms}.
\end{equation}

We note that the leading correction to asymptotic value of the exponent is 
proportional to $2 - {\tilde d}_b$. The next correction term is of order
$1/\log b$, and is proportional to $2 - D_b$, where $D_b = 2 
-{\bar \epsilon}$ is 
fractal dimension. These are like the $\epsilon $-expansions, 
except that  there are several inequivalent definitions of 
dimension  for fractals.  Thus there are several  $\epsilon$'s, and the 
exponents on fractals may require a 
multi-variable $\epsilon$-expansion. Interestingly, at higher orders, 
corrections to scaling to the finite-size scaling functions $f(x), g(x)$
etc. would give corrections to the exponents of the type $1/b^{\Delta}$.
These are of the type $ \exp ( - \Delta/{\bar \epsilon})$, and such 
correction
terms are not calculable within the conventional  
$\epsilon$-expansions framework.

The reason the critical exponents in the large $b$ limit do not tend to 
the Euclidean value may be understood as a crossover effect. For large 
$b$, the space looks Euclidean at length scales smaller than $b$, and the 
effective polymer exponents for $\epsilon > 1/b^{1/\nu}$ would be near the 
Euclidean values. However, for $\epsilon \ll 1/b^{1/\nu}$, the polymer has 
to go through the constrictions, and the asymptotic value of exponents
for large polymers can, and do, take different values.

\section{THE COLLAPSE TRANSITION IN  POLYMERS WITH SELF-ATTRACTION}

We have seen that the qualitative features of the behavior of linear
polymers with no interaction other than the excluded volume interaction is
well-modelled by SAWs on fractals.  Now we will like to show that
polymers on fractals can also be used to understand more complicated
behavior of polymers like the transition from the swollen state to compact
globular state transition that is observed in dilute polymer solutions as
the temperature is lowered \cite{Tanaka}. 

In order to model this, we have to include the attractive interaction
between different parts of the polymer when they are close by, but not
overlapping. The simplest lattice model for this phenomena associates an
energy $- E_u$ for each pair of nearest-neighbor lattice points occupied
by the polymer that are not consecutive along the walk. The equilibrium
properties of such a self-attracting SAW can be described using the grand
partition function $G(x,u)$;

\begin{equation}
G(x,u) = \sum_{N,N_u}\Omega(N,N_u) x^N u^{N_u},
\label{eq66}
\end{equation}
where $\Omega(N,N_u)$ is the number of different configurations per site
of a self-avoiding polygon of $N$ steps and energy $-N_u E_u$, $u = \exp(
\beta E_u)$, $x$ is the fugacity per step of the chain, and $\beta $ is
the inverse temperature. 

 For small $\beta$, the effect of $E_u$ can be ignored, and the typical
size of polymer varies as $N^{\nu}$, where the exponent $\nu$ takes the
value for SAWs. This is called the swollen state of the polymer. For
large $\beta$, the polymer folds up like a  tangled ball of yarn, in order 
to
minimize the energy $-N_u E_u$. In this phase, called the collapsed phase,
the typical size of polymer varies as $N^{1/d}$ in $d$ dimensions.

In the limit when the number $N$ of monomer units goes to infinity, there
is sharp transition from the swollen to collapsed phase at a critical
value of $u$. This transition is described as a critical phenomena
analogous to a tricritical point for magnetic system \cite{degennes}. For
large  polymers, the average gyration radius $R$ at the
transition behaves as $R_N \sim N^{\nu_{\theta}}$ where the exponent
$\nu_{\theta}$ is intermediate between the value $\nu$ for swollen state
and the value $\nu_c = 1/ D$ for compact globule on a lattice of
fractal dimension $ D$.

\subsection{Self-interacting polymer on the $3$-simplex lattice}

The properties of a polymer chain with self-attraction on  a fractal were first 
studied by Klein and Seitz  \cite{DW}. They used the self-avoiding walks 
on the Sierpinski gasket, which is the $b=2$ member of the Given-Mandelbrot 
family. We consider below the case of $3$-simplex, which is somewhat simpler
to treat.

\begin{figure}[ht]
\begin{center}
\includegraphics[width=3.0in,height=1.0in]{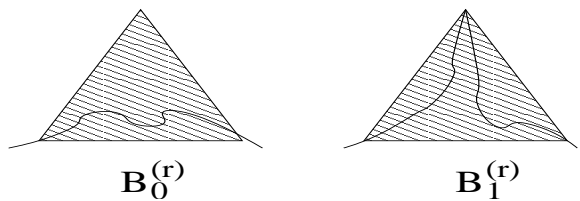}
\caption{Restricted partition functions for the self-attracting walk for 
the $3$-simplex.} 
\label{klein}
\end{center}
\end{figure}

To calculate  $G(x,u)$ for the $3$-simplex lattice,  we  
define restricted partition functions $B_0^{(r)}$ and $B_1^{(r)}$,  for 
walks that cross the $r$-th order triangle, as shown in   
Fig. \ref{klein}. $B_0^{(r)}$ is the sum of weight of all walks that enter an $r$-th order 
triangle of the $3$-simplex from one corner and leave from another corner, but do not 
visit the third corner. $B_1^{(r)}$ is the sum of weights of walks that
enter and leave the $r$-order triangle, and also visit the third corner of 
the triangle.  Then it is easy to see that these weights satisfy the 
recursion equations
\begin{equation}
B_0^{(r+1)} = [ B_0 + B_1 ]^2 +  B_0 [ B_0^2 + 2 B_0 B_1  +  B_1^2 u ],
\label{eq53}
\end{equation}
\begin{equation}
B_1^{(r+1)} = B_1 [ B_0^2 + 2 B_0 B_1  + B_1^2 u ].  
\end{equation}

The generating function for all loops is given by the formula
\begin{equation}
P(x,u) = \sum_{r=1}^{\infty} [ B_0^{(r)} + B_1^{(r)}]^3 3^{-r}.
\end{equation}
We start with the initial weights
\begin{equation}
B_0^{(1)} =0; ~~ B_1^{(1)} = x.
\end{equation}
These variables tend to the  trivial fixed point $B_0^* = B_1^* = 0$, if 
the starting value of  $x$ is less than  a critical value $x_c(u)$, and to the 
fixed point $B_0^* = B_1^* = \infty$, if $x > x_c(u)$.  For $x = x_c(u)$,  $B_1^{(r)} $ 
tends to zero for large $r$, and the Eq. (\ref{eq53}) reduces to Eq.(\ref{eq8}). 
The critical properties of the chain are continuous functions 
of $u$, and there is no phase transition as a function of $u$.

We note that here the recursion equations involve the interaction 
parameter $u$. This complicates the analysis. Consider a modified problem 
where the interaction $-U$ occurs only between the nearest neighbor bonds 
in the same $2$-nd order triangle, and not otherwise. One would expect the 
qualitative behavior of the system very similar, but now, the recursion 
equations do not involve $u$. In fact, they are the same as the case 
without self-attraction [Eq.(\ref{eq8})]. This observation is helpful in 
studying the collapse on other fractals.

\subsection{ Collapse transition on $n >3 $ -simplex lattices}
 
The collapse transition on the $4$-simplex lattice was first studied in
\cite{dv}. We restrict the attractive interaction to nearest neighbors 
within  the same  2nd order simplex. Then, the recursion equations Eq. 
(\ref{eq26}-\ref{eq27}) describe the collapse transition also, if we use 
the starting weights

\begin{equation}
A^{(2)} = x^2 + 2x^3 u + 2x^4u^3, \qquad {\rm and} \qquad B^{(2)} = 
x^4u^4.
\end{equation}

\begin{figure}
\begin{center}
\includegraphics[width=6cm,angle=0]{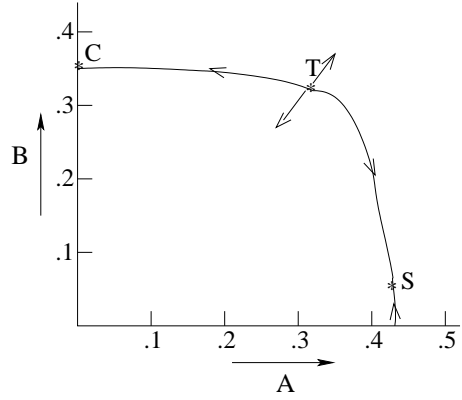}
\caption{Different fixed points for the $4$-simplex lattice. The 
non-trivial fixed points labelled $C$,$T$ and $S$ correspond to the 
collapsed, the tricritical $\theta$-point, and the swollen state. } 
\label{4simplexfp} \end{center}
\end{figure}

The renormalization group flows of the variables $A$ and $B$ in this case 
is shown in Fig. \ref{4simplexfp}. There is the trivial fixed point 
$A^* = B^* = 0$, which 
is  reached if  $x < x_c(u)$. If $x > x_c(u)$, the fixed point $A^* = B^* 
= \infty$ is reached. The two-dimensional space of possible initial 
conditions $( A^{(1)}, B^{(1)})$ is divided into the basins of attraction 
of these two fixed points. The common boundary of these basins is a line.
This line is an invariant sub-manifold of the renormalization flows (i.e. 
points starting on the line remain on the line).  On this line we have   
three  fixed points : 
\begin{enumerate}
\item The fixed point $(0.4264, .04998)$ corresponds to the swollen state. 
This is reached for $u < u_c \simeq 3.106074$ and $x = x_c(u)$. This is 
denoted by S in Fig. \ref{4simplexfp}.  If we start at a point near this 
fixed point on the critical 
line, the renormalization flows are towards this fixed point. 

\item 
The fixed point $(A^*, B^*) = (0,22^{-1/3}) $ is reached when $u >
u_c$. It gives one relevant eigenvalues $\lambda = 4$
corresponding to $\nu_c = 1/D_4 = 1/2$. In  this  phase, the polymer
fills the available space with a  finite density of monomers. This fixed 
point is denoted by C in Fig. \ref{4simplexfp}. This is also attractive 
for  points starting nearby on the critical line.

\item The fixed point $(1/3, 1/3)$, denoted by T in Fig. \ref{4simplexfp}, 
is obtained for $u=u_c$. On the 
critical line $x = x_c(u)$, this unstable fixed point lies between the 
fixed points corresponding to the swollen and the collapsed phases. The 
matrix corresponding to linearized renormalization transformation at this 
point has two
eigenvalues greater than one; $\lambda_1 = 3.7037$ and $\lambda_2 = 2.2222
$.   The third relevant eigenvalue corresponds to the renormalization 
of fugacity of end points ( $\lambda_+$ in Eq. (18)). Thus this is a 
tricritical point. at this point, $\nu_c = \ln 2 /\ln \lambda_1 
= 0.5239$.  The
singular part of the free energy varies as $(u-u_c)^{2-\alpha}$ with 
\begin{equation}
\alpha = 2 - \frac{\ln \lambda_1}{\ln \lambda_2} = 0.36027.
\end{equation}
\end{enumerate}

One can similarly study the $n=5$ simplex lattice \cite{Singh1},  with 
the interactions are confined to the internal
bonds of the second order graph. It is found that just like the $n=3$
case, there is no collapsed phase. One can understand this by noting that
the ground state configuration corresponds to a walk that visits all the
$5$ sites of the $2$nd order subgraph.  There are many
configurations corresponding to the minimum energy, and most of these are
extended. Even when there is attractive interaction between all pairs 
of neighbors, the energy-cost of pulling a polymer  confined to an
$r$-th order subgraph  to something that is spread over a subgraph of 
order $(r+1)$ is finite. As there is a large entropy associated with the
place where this break occurs, this makes the collapsed phase unstable to 
such breaks, and brings the collapse transition temperature to zero.

In a similar way, we can study the $6$-simplex. Here we have coupled 
polynomial recursion equations of degree $6$ for the three variables 
$A^{(r)}, B^{(r)}$ and $C^{(r)}$.  For the details of the recursion 
equations, see  \cite{Singh1}. Again, we have two trivial attractive 
fixed points corresponding to $(A^*, B^*, C^*)$ equal to $(0,0,0)$ or 
$(\infty,\infty,\infty)$. There is a two-dimensional critical manifold 
that marks the boundary of the basins of attraction of these fixed 
points.  For the renormalization flows on this  two-dimensional  critical 
surface, there are 
two attractive fixed points:

The fixed point
$(A^*,B^*,C^*) = (0.262352, 0.017586, 0.000701)$ corresponds to the
swollen state as discussed in Sec. 4.1. 

The fixed point $(A^*,B^*,C^*) = (0,0,0.071329)$ is reached for all $u>u_c
(= 3.4999847)$ at $x=x_c$. The largest eigenvalue for this case is
$\lambda = 6$ corresponding $\nu_c = 1/{D_6} = 0.3869$. This describes 
the collapsed phase of the polymers. 

The common boundary of  basins of attraction of these two fixed points is  
a $1$- dimensional line, which  
is also an invariant submanifold for the renormalization flows. This line 
has one attractive fixed point:

The fixed point (0.12949,0.09572,0.05344) is obtained for $u=u_c$ and
$x=x_c$ and has two eigenvalues greater than one; $\lambda_1 = 5.4492 \;
{\rm and} \; \lambda_2 = 1.9049$. This is a tricritical point with
exponent $\nu_t = 0.4088$ and $\alpha = 2 - \ln \lambda_1/\lambda_2 =
0.6309$.

The crossover exponent $\phi$ at the tricritical point is $\phi =
\nu_t/\nu_{th}= 0.3801$, where the exponent $\nu_{th}$ controls the
divergence of thermal correlation and is defined as $\nu_{th} = \ln 2/\ln
\lambda_2 = 1.0755$.

There are two other fixed points (0.254037, 0.022159, 0.07098) and
(0.2000, 0.0666, 0.0666)  on the line separating the basins of attraction
of the fixed points corresponding to the collapsed and the swollen phases.
These are purely repulsive, and cannot be reached starting with our choice
of initial condition. These correspond to higher order multicritical
points( tetra-critical).

  We note that the collapse transition on fractal lattices
corresponds to a new fixed point, intermediate between swollen (SAW) phase
and the collapsed phase and cannot be viewed as a perturbation of the
Gaussian fixed point describing random walks.

\subsection{Some unusual phases }

The nature of possible collapsed phases of linear polymers on fractals
with a finite ramification number depends strongly on the geometrical
constraints imposed by the connectivity properties of the the fractal,
much more so than in the extended phase.

As we already noted, for the $n$-simplex lattice with $n$ odd, a collapsed
phase with a finite density of the polymer is not found. For the modified
rectangular lattice, the behavior of linear polymer with self-attraction
was studied in \cite{dv}.  Here the number of variables needed to describe
the polymer with self-attraction becomes rather large. For example, to
describe closed loops it is necessary to introduce additional weights
$A_1^{(r)}, ~ B_1^{(r)}, A_2^{(r)}, \ldots$ etc., where the subscript 1(2)
indicates the number of extra corner sites of the $r^{th}$ -order are
visited, a total of eleven variables. To describe open chains, we would
need 17 additional variables, making a total of 28 variables- a rather
formidable number.  However, as in case of  $n$-simplex lattices
many of these variables are irrelevant and may be set equal to zero. 
Equivalently, we study collapse when the attractive interaction is 
restricted to the first order rectangles. If we
restrict ourselves to the analysis of closed polygons, we require only the
five variables defined earlier in section 4.3.

Interestingly, one finds three different phases of the polymer, depending
on the value of the attraction parameter $u$. For small $u < u_{c1} =
3.2023$, the polymer exists in the swollen phase, with typical size $R
\sim N^{\nu}$, and $\nu = 0.6650$. For large $u > u_{c2} = 3.2341$, it
exists in a compact phase of finite density, with $R \sim N^{1/2}$.
However, between the swollen and compact phases, for $u_{c1} < u < u_{c2}$,
one finds a `rod-like phase', where the shape of the polymer is highly
anisotropic.  In the x-direction, the average extent of polymer increases
as $N$, and in the other direction it remains finite. At $u =u_{c1}$,
there is a non-trivial  fixed point of the recursion equations of period 
$2$.
Linear analysis of the renormalization equation about this fixed point
gives $\nu =0.80503$. At $u= u_{c2}$, one gets $\nu = 1/2$, same as in the
collapsed phase, with possible logarithmic corrections.  There is a
logarithmic singularity in the specific heat also. The reason why the
rod-like phase is found in this case is not clear : obviously the
anisotropy of the lattice is responsible for it, but it does not matter so
much in other( swollen or collapsed ) phases, where the anisotropy
is of the usual type ( the ratio of average diameter of an $N$-stepped 
walks in the $x$- and $y$- directions is a finite number).

Another different type of phase, labelled `semi-compact' phase was found
by Knezevic and Vannimenus in their analysis of the collapse transition on
the $HB(3,3)$ fractal \cite{MJ}.  In this case, the connectivity of the
graph is such that a linear polymer cannot fill the available space with a
finite density. For large value of the attraction strength $u$ the polymer
shrinks into a ``semicompact'' state. In this phase, the average monomer
density tends to zero for large polymers.

The recursion relations found for restricted generating functions $A^{(r)}$
and $B^{(r)}$ where $A^{(r)}$ counts the number of configurations when the
polymer goes once through the $r^{th}$-order gasket while $B^{(r)}$ counts the
configurations where the polymer goes twice through the gasket, are rather
complicated \cite{MJ}, but for the large $u$ regime, the equations are
dominated by only a single term in each polynomial. For large order of
iteration, the recursion equations in this phase may be approximated by
\begin{equation} 
A^{(r+1)} \simeq 320 A^3 B^6; \qquad  B^{(r+1)} \simeq 4308 A^2 B^8.  
\end{equation}

These can be linearized by taking the logarithm of both sides, which shows
that for large $r$, $\log A^{(r)}$ and $\log B^{(r)}$ vary as
$\lambda_{+}^r$, where $\lambda_{\pm} = (11 \pm \sqrt{73})/2$. They will
both diverge to $+\infty$, or to $-\infty$ depending upon the coefficient of
proportionality being positive or negative ( both are of same sign for the
largest eigenvalue). Thus, this constant must be proportional to distance
from the critical line $x -x_c(u)$. If $x_c - x = \delta \ll x_c$, number
of iterations before the deviation becomes of order 1 is 
\begin{equation}
r_0 \approx \log (1/\delta)/ \log \lambda_{+}. 
\end{equation} As the size of
the polymer is approximately $ 2^{r_0} \sim (1/\delta)^{\nu'}$, this
corresponds to $\nu' = 0.48195$. The fractal dimension of the chain in
this phase is, $1/\nu' = 2.07491 $ a value just a bit less than
the fractal dimension of the lattice $ D = 2.09590$. For $x = x_c$,
$ A^{(r)}$ decreases to zero as $\exp( - a \lambda_{-}^r)$ where $a $ is
some constant, and $B^{(r)}$ increases to infinity as an inverse power of
$A^{(r)}$.

\begin{figure}[ht]
\includegraphics[width=5.0in,height=1.5in]{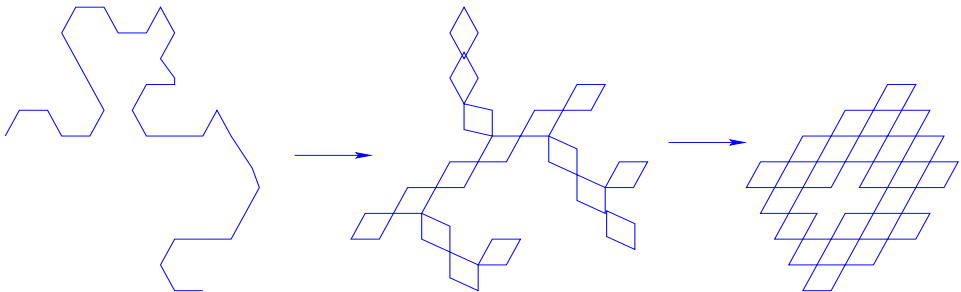}
\caption{ Schematic representation of the different phases of a linear 
chain with self-attraction on the Sierpinski gasket with at most two 
visits per site allowed: (a) the swollen 
phase (b) branched polymer phase with branches made of the doubled-up 
chain (c) the collapsed phase} 
\label{orlandini}
\end{figure}

In all the cases discussed so far a restriction on the walks that a
lattice bond or a lattice site visited once cannot be visited again has
been imposed. One can relax this condition and allow the walk to revisit a
site already visited so long as no bond is traversed twice. A consistent
set of visited bonds is called a trail, if different sequences in which
the same set of bonds bonds may be visited are considered equivalent
\cite{Malaki}. Orlandini et al \cite{Orlandini} considered self-attracting
SAWs on a 2d Sierpinski gasket, with this constraint, and showed that this
model displays a new multicritical point corresponding to a collapse from
linear into branched polymer in which the polymer becomes a randomly
doubled-up chain ( see fig. \ref{orlandini}), which is followed by a
further transition into compact globule. In Fig. \ref{orlandini}, we have
drawn schematically what the different phases look like ( drawn here for
the Euclidean 2-dimensional space).  The universality class of the two new
multicritical points corresponding to the phase transition points is
different from the tricritical points discussed so far.

\section{BRANCHED POLYMERS}

The models of branched polymers is related to  other important problems in 
statistical mechanics, such as lattice animals,
percolation \cite{Harris} and the Lee-Yang edge singularity \cite{GN}. The
study of branched polymers on fractal lattices are therefore very
instructive as one can analyze the system in detail. The number of
different branched polymers made of $N$ monomers and their average
gyration radius are expected to grow as $\mu^NN^{-\theta}$ and $N^{\nu}$,
respectively for large $N$; $\theta$ and $\nu$ are critical exponents,
whose values are different from those from that of linear polymers. If we allow
loops in the cluster, we get the model of lattice animals, which
corresponds to the $p \rightarrow 0$ limit of the percolation problem.
These are known to be in the same universality class as branched polymers.
If in addition to the excluded volume interaction, one has an attractive
nearest-neighbor interaction between the monomers, the branched polymers
can undergo a collapse transition, just like the linear polymers. Near the
collapse point, the free energy per monomer shows singular critical
behavior.

Knezevic and Vannimenus realized that real-space renormalization method
for studying linear polymers on fractals can be extended directly to the
case of branched polymers \cite{kv1,kv2}. They considered asymptotic
properties of branched polymers with attractive self-interaction on
fractal lattices, restricting the attractive interactions to bonds within
first order units of the fractal lattices. We summarize their findings
here.

\subsection{The  $3$-simplex lattice}

\begin{figure}[ht]
\includegraphics[width=6.0in,height=2.1in]{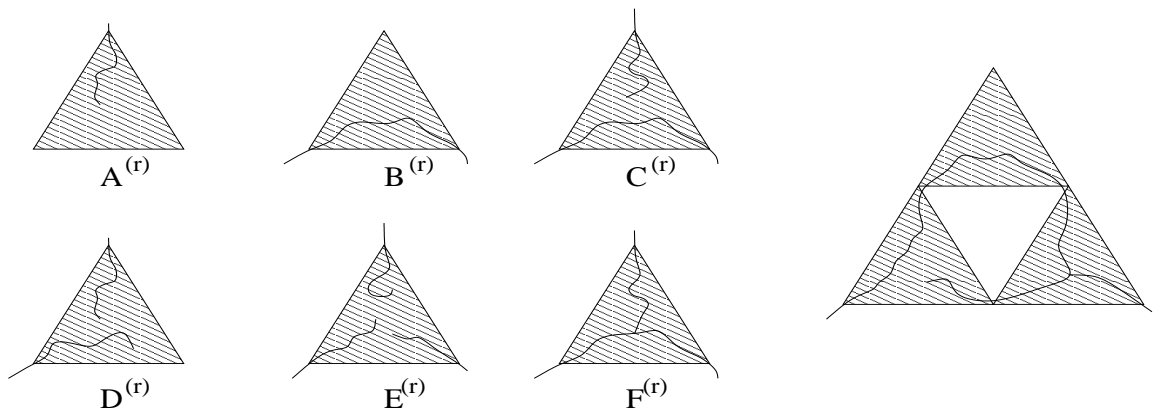}
\caption{ Diagrams representing the six restricted generating functions for branched polymers
on the two dimensional Sierpinski gasket. $C^{(r)}$ corresponds, for 
instance, to configurations
where a part of the polymer joins two vertices of an $r^{th}$ order triangle while one of 
its ends penetrates through the third vertex. The  diagram on the right 
shows a term 
$B^{(r)}C^{(r)}F^{(r)}$ contributes to $B^{(r+1)}$.}
\label{bp3simplex}
\end{figure}

The macroscopic thermodynamic quantities of interest can be obtained from   
the grand-partition function $G(x,u)$,
whose definition is same as Eq.(\ref{eq66}), with $\Omega(N,N_u)$ now
defined as the average number of unrooted branched polymers with $N$ 
bonds,
and $N_u$ nearest neighbor bonds, the average being taken over different
positions of the polymer on the fractal. This can be determined in terms 
of 
the $r$-th order restricted partition functions defined in Fig. \ref{bp3simplex}.

The closed set of recursion equations involve six restricted generating
functions as described in Fig \ref{bp3simplex} are easily  written down
\cite{kv1}: 
\begin{eqnarray} 
A^{(r+1)} & = & A[1+2B+2B^2] + 2B^2C + F[B^2+A^2+2BD], \nonumber \\ 
B^{(r+1)} & = & B^2 + B^3 + F[4BC + 2AB] + F^2[B+D], 
\end{eqnarray} 
and similar equations for other variables. $G(x,u)$ can be seen as a 
sum of terms of the same general form as Eq. (\ref{eq5}) and 
(\ref{eq6}).  The  singular behavior of the sum can be analyzed by looking
at the fixed points of the recursion equations. There are two trivial 
fixed points with all variables  zero or infinite. For any given value of 
$u$, we have to tune the initial value of $x$ to a critical value $x_c(u)$ 
to get a non-trivial fixed point.

The analysis of the recursion equations lead to following three different
nontrivial fixed points: 
\begin{enumerate} 
\item For all $u< u_c = 5.5$, and $x = x_c(u)$, the recursion equations 
lead to a non-trivial fixed point $(A^*,B^*, \ldots)$ corresponding to the 
swollen state with large scale properties same as for the
random-animal problem ($u$=1).  All the fixed point values $A^*, 
B^*,\ldots$ at this fixed point are non-zero. Linear analysis of the 
recursion equations about this fixed point shows that the  
generating function $G(x,u)$ has a 
power-law singularity when $x=x_c(u)$ and the critical exponents are $\nu =
0.71655$ and $\theta = 0.5328$. 
\item For $u>u_c$, and $x = x_c(u)$,  the polymer is in the 
collapsed phase. The corresponding  fixed point has  $A^* = B^* =D^* 
=0$. $F^{(r)}$ tends to zero, and $C^{(r)} F^{(r)}$ and $E^{(r)} 
{F^{(r)}}^2$ tend to a finite limiting value.  The largest value 
of the linearized renormalization transformation matrix is $3$, corresponding to 
$\nu_c = 1/ D_3 = 0.63093$. This dense phase of 
branched polymers is in the same universality class as the spanning trees.
For the spanning trees, one can also define a chemical distance exponent
$z$, by the relation $ \ell \sim R^z$, where $\ell$ is the distance along 
bonds between two randomly picked points on the polymer, and $R$ is the 
euclidean distance. This was calculated in \cite{ddad}, and shown that $z 
= \log [(20 +\sqrt{205})/15]/ \log 2 \simeq 1.1939$.
    
\item 
The collapse transition occurs at $u=u_c$. At this fixed point, all the values 
$A^*,B^*,\ldots$ are non-zero. This is a tricritical  fixed point, with 
two eigenvalues larger than one.  The exponent $\nu_t
= 0.63250$ is very close to $\nu_c$. The exponent $\alpha$ is negative,
$\alpha = -4.0269$, showing that the singularity in the specific heat at 
$u=u_c$ is very weak.  
\end{enumerate}

The closeness of $\nu_t$ to $\nu_c$ has also been found on square lattice
where an accurate transfer - matrix study \cite{Derrida} of the collapse
of branched polymer gives $\nu_t = 0.509 \pm 0.003$ which is very close to the compact
value 1/2. This suggests that the phenomenon is not accidental and should 
have  a more general explanation.

The problem of linear polymers is recovered if all terms containing the
functions $E$ and $F$ one suppressed in Eqs.(7.1) and (7.2). The truncated
equations have only one fixed point, corresponding to the swollen phase
with $\nu = 0.7986$ and there is no collapsed phase (see Section 3).

\subsection{The 4-simplex lattice }

The closed recursion equations in this case involve eleven restricted
generating functions. The number of polymer configurations to consider is
large, and Knezevic and Vannimenus \cite{kv2} used computer enumeration to
sort them out. We omit the details.

 The analysis found  three non-trivial  fixed points that describe the 
large-scale behavior of self-interacting
branched polymers. 
\begin{enumerate} 
\item For $u<u_c \approx 2$ the
random - animal fixed point corresponding to the swollen phase of the
polymer is reached. Linearizing around the fixed point one finds only one
relevant eigenvalues $\lambda_1 = 3.14069$. With this eigenvalue one finds
$\nu = \log 2/ \log \lambda_1 = 0.60566, \; {\rm and} \; \theta = 
0.75667$.
\item For $u>u_c$ one gets a fixed point corresponding to the collapsed
state in which polymer occupies all vertices of the lattice. The relevant
eigenvalue found in this case is $\lambda = 4 $ which gives $\nu_c =
1/2$. 
\item For $u=u_c$ the fixed point reached represents a tricritical
point as it has two eigenvalues greater than $1$: $\lambda_1 = 3.94050$ 
and $\lambda_2 =
1.32094$. The exponent $\nu_t = \log 2/ \log \lambda_1 =
0.50546$. This value is very close to the value $\nu_c = 1/2$ of the
collapsed phase. The exponent $\alpha = -2.8267$ is negative but less
negative than for the 3-simplex lattice.  \end{enumerate}

\subsection{Other fractals}

Knezevic and Vannimenus also studied the branched polymer problem on the
$GM_3$ fractal \cite{kv1,kv2}.  In this case, they found the unexpected
result that unlike the case on the $GM_2$ fractal ( for which the behavior
is the same as on the $n=3$ simplex), for the $b=3$ case, the number of
animals of size $n$ grows as $\mu^n \exp( \kappa n^{\psi})$, where $0 < 
\psi
<1$. This corresponds to an essential singularity in the generating 
function of branched polymers $G(x,u=1) \sim \exp(\frac{a}{|x_c 
-x|^\rho})$,
with $\rho =\psi /(1 - \psi)$. 

The analysis of Knezevic and Vannimenus  has recently been extended to all $b$, 
and one finds
that for $b \geq 3$, the average number of animals per site behaves as
$\mu(b)^n \exp( \kappa n^{\psi})$, where the  values of the the
singularity exponent $\psi$, and the size exponent $\nu$ can be determined
exactly \cite{ddgiven}.

\section{SURFACE ADSORPTION}

It is well known that a long flexible polymer in a good solvent can form a
self-similar adsorbed layer near an attractive wall at the critical
temperature $T_a$. Using the correspondence between an adsorbed polymer 
chain and
the model of ferromagnets with  $n$-vector spins in the limit $n 
\rightarrow 0$  with a free surface, it has been 
shown that
the adsorption point $T_a$ corresponds to a tricritical point and in its
proximity a crossover regime is observed. In particular, the mean number
of monomers, $M$, at the surface is shown to behave as \cite{bell}

\begin{eqnarray}
M & = &  (T_a-T)^{(1/\phi - 1)},  ~~~~~~~~{\rm ~for~} T < T_a; \nonumber  
\\
& =& N^{\phi}, ~~~~~~~~~~ {\rm ~for~} T = T_a;  \nonumber \\
& = & (T-T_a)^{-1},   ~~~~~~~~~{\rm ~for~} T > T_a.
\end{eqnarray} 	 

A good model for this phenomenon is a SAW on some lattice with an
absorbing surface (boundary). Every site on surface visited by the polymer
contributes an energy $-E_s$. This model has
widely been studied on various lattices and via a number of techniques
that include exact enumeration \cite{rajesh,rajesh2}, Monte Carlo 
\cite{grass},
transfer matrix \cite{veal}, renormalization group \cite{kumar1} etc. For
a 2-d Euclidean lattice, exact value of $\phi$ found from conformal field
theory is 1/2 \cite{burkhardt}.

Bouchaud and Vannimenus \cite{bouchaud} were the first to apply RSRG 
techniques on fractals to study the linear polymers near an attractive 
substrate. They  showed that the known phenomenology of the adsorption 
-desorption transition is well-reproduced on fractals, and  different 
critical exponents can be evaluated exactly.  The 
values of the exponent $\phi$ for $HB(2,2)$ and $HB(2,3)$ fractals were 
found to be 
0.5915 and 0.7481 respectively. They also showed that for a container
of fractal dimension $D$ and adsorbing surface $d_s$, $\phi$ has lower and upper bounds;
\begin{equation} 	 
1-(D-d_s)\nu \leq \phi \leq \frac{d_s}{D}.
\end{equation}

\subsection{Surface adsorption in polymers with self-attraction}

Polymer chains with self attraction near an attractive surface can exhibit
a rich variety of phases, characterized by many different universality domain
of critical behavior [57]. This is due to competition between the two 
interactions. At the intersection of  two tricritical surfaces, one 
corresponding to the 
$\theta-$transition and another to the adsorption-desorption transion, one 
can expect higher 
order critical points.
Understanding of the different phases that are possible and understanding 
and classifying the multicritical points that can exist appear difficult
on standard Euclidean lattices[47]. It is here fractal lattices have been 
particularly helpful [58]. The values of critical exponents found are, of
course, different for different fractals, but the general features
of the phase-diagrams remain the same. Investigating the problem on
fractals helps us understanding the problem in real 
experimental systems.

We consider a linear polymer chain on a truncated $n$-simplex lattice and
make one boundary surface of it attractive.  We treat one of the edges of
the fractal container as an attractive surface and associate an energy
$-E_s < 0$ with each site on it occupied by the polymer, and an energy
$E_t > 0$, with each occupied site that is at a distance $1$ from the
occupied surface i.e. on the layer adjacent to the surface, and an energy 
$-E_u$ for  nearest-neighbor bond between monomers not consecutive 
along the chain.  

Then, to  each $N$-step walk having 
$N_s$
steps along the surface, $N_t$ steps lying on the layer adjacent to the
surface and $N_u$ number of nearest neighbors a weight 
$x^{N}\omega^{N_s}t^{N_t}u^{N_u}$ is assigned, where $\omega=\exp(\beta 
E_s)$, and $t = \exp(- \beta E_t)$.   The grand partition 
function for this system is given by                
\begin{equation}
G(x, \omega, t, u)= \sum_{N, N_s, N_t, N_u}\Omega(N, N_s, N_t, N_u) x^{N}\omega^{N_s}
t^{N_t}u^{N_u},
\label{eq85}
\end{equation}
where $\Omega(N, N_s, N_t, N_u)$ is the number of different configurations
per site of a $SAW$ of $N$ steps, rooted at a specified site on the 
attractive surface, with given values of $N_s, N_t$ and $N_u$.
We may put $E_{t}=0$ (or $t=1$) for simplicity, but a non-zero value
seems to give rise to interesting behavior, and is also important to 
many physically realizable cases. The attractive interaction between monomers 
is restricted, as in preceding sections, to bonds within the lowest order 
subgraphs of the fractal lattices.

As shown in section 6.2, polymers in the n-simplex container show a 
collapsed phase only for even values of n. Thus, the 4- and 5-simplex
lattices exhibit contrasting behavior and represent 
two different scenario which may arise in real systems. In the case 
of 5-simplex lattice, there is no collapse transition possible in the 
bulk, but its 4-simplex surface can show a collapsed phase. In the case 
of 4-simplex lattice, on the other hand, there is a collapsed globule
phase in the bulk but no collapse possible in the surface-adsorbed polymer.
We can therefore have situations in which the bulk acts as a poor-solvent
medium, while the surface acts as a good solvent medium or the 
opposite case of surface being poor-solvent medium and the bulk good solvent.
However, the situation in which both bulk and surface show collapsed phases can 
not be modelled by polymers on an $n$-simplex fractal.  
 
\subsection{The $4$-simplex lattice}

\begin{figure}[ht]
\begin{center}
\includegraphics[width=4.0in,height=2.0in]{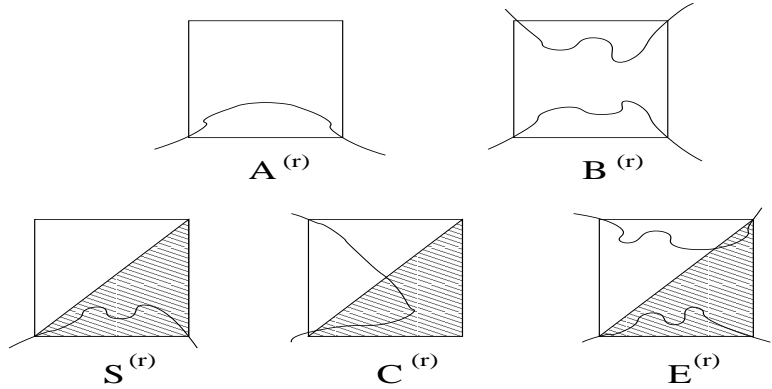}
\caption{The restricted partition   functions for the  $r$th order 
$4$-simplex with attractive interactions at one boundary 
surface. The internal structure of the $4$-simplex is not 
shown. The shaded triangle reprsents the attractive surface.}
\label{fig91}
\end{center}
\end{figure}

\begin{figure}[ht]
\begin{center}
\includegraphics[width=2.5in,height=2.0in]{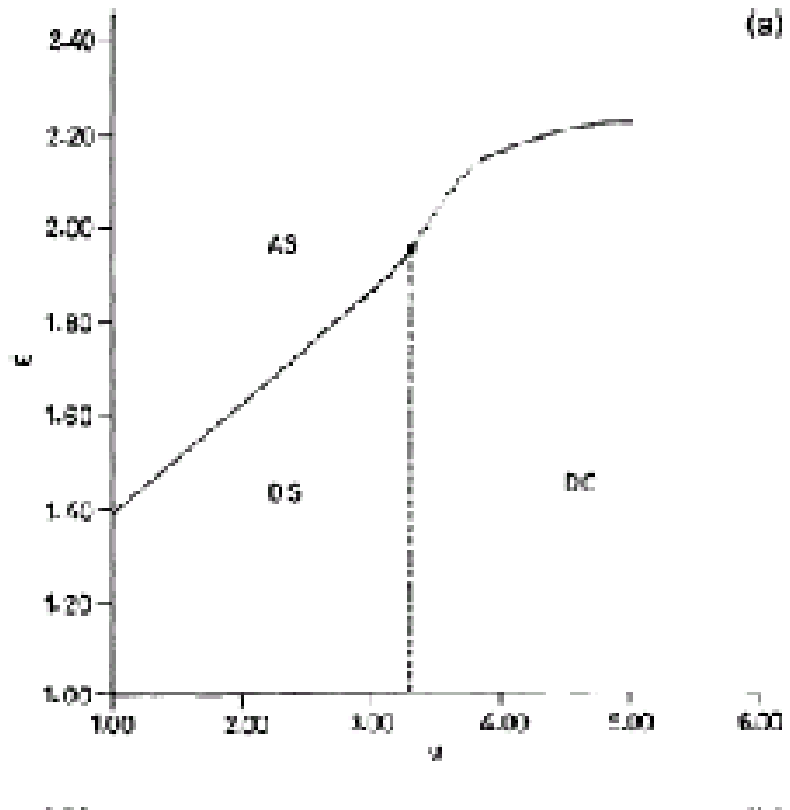}
\includegraphics[width=2.5in,height=2.0in]{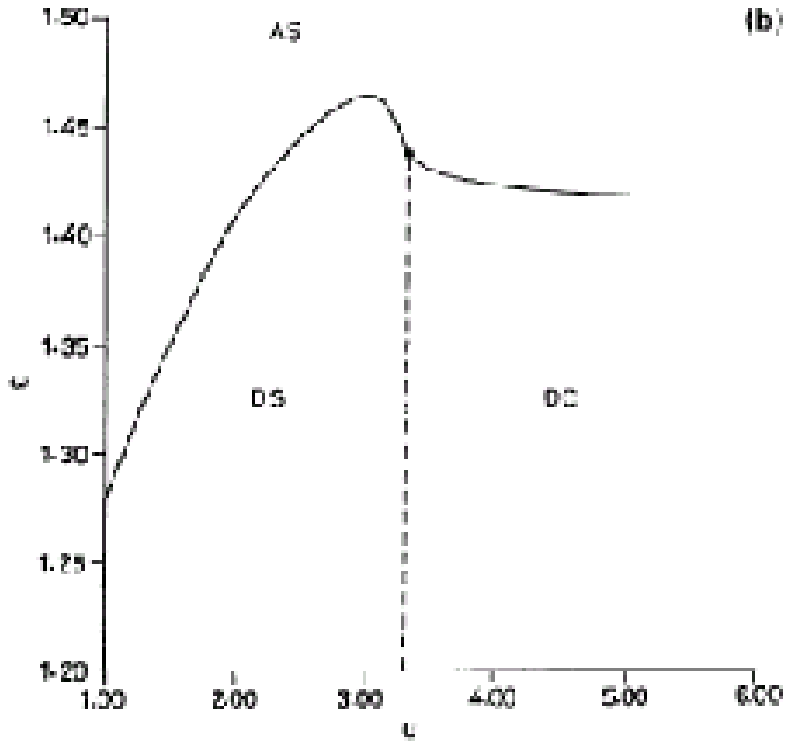}
\caption{ Sections of the $\omega-u-t$ phase diagram for two different
values of $t$ for the 4-simplex lattice: (a) t= 0.2 (b) t= 0.5.  Regions
marked by AS, DS and DC represent, respectively, the adsorbed polymer in
swollen state, desorbed polymer in swollen and collapsed (globular) state.
The collapse transition between the DC and DS phases is denoted by the
dashed line. The special adsorption line is indicated by full line and
part of it by dotted line. The point where $\theta$-line meets with the
adsorption line is a multicritical point. The dotted part of the
adsorption line indicates the region of coexistence of adsorbed SAW and
the (bulk) globule phase.}
\label{fig92}
\end{center}
\end{figure}

For the 4-simplex lattice the grand partition function of Eq. (\ref{eq85}) 
can be written in terms of the five restricted partition functions shown
in Fig 17. The shaded regions in the figure represent the surface. Out 
of five configurations, two ($A^{(r)}$ and $B^{(r)}$) represent the sum
of weights of configurations of the polymer chain within one r-th order 
subgraph away from the surface, and the remaining three ($S^{(r)}, C^{(r)}$
, and $E^{(r)}$) represent the surface functions. The recursion
relations for these restricted partition functions can easily be written
(see [58] for details).  The equations for $A^{(r+1)}$ and $B^{(r+1)}$
do not get affected by the presence of the surface interactions. 

In this case, we have three phases possible: the desorbed swollen(DS), the
desorbed collapsed (DC) and the adsorbed swollen (AS) phases. The fixed points
corresponding to the desorbed phases have $ S^*= C^*= E^*= 0$, and
$A^*,B^*$ equal to the value for the $4$-simplex at the S or C fixed point
( section 6.2). The fixed points corresponding to the polymer adsorbed on
the surface, on the other hand, have $A^*= B^* = C^*= E^*= 0$ and $S^*$
equal to the value $\mu^{-1}$ corresponding to the swollen phase fixed
point on the $3$-simplex surface ( section 3.1).  The phase boundaries,
determined numerically by finding the basins of attraction of these fixed
points are shown in Fig. \ref{fig92}.

If we start at any point on the boundary between the DS and AS phases,
under renormalization, we flow to the symmetrical fixed point $ S^* = C^*
= A^*, B^* = E^*$, with $A^*, B^*$ having the value corresponding to the
point $S$ in Fig. \ref{4simplexfp}.  Linearizing the recursion equations
near this fixed point two eigenvalues $\lambda= 2.7965$ (corresponding to
the swollen bulk state) and $\lambda_{\phi}= 2.1583$ greater than one
[58] are found.  This point is the expected symmetrical special absorption point
which describes the polymer at the desorption transition. A simple
calculation gives the crossover exponent $\phi= 0.7481$ and $\alpha=
0.6653$. In Fig 18, the full line shows intersection of this surface with
the surface $t =$ constant.

The bulk transition between DC and DS phases is described by the fixed 
point $S^* = C^* = E^* = 0$, and $A^*$ and $B^*$ have the value (1/3,1/3)
corresponding to the point T in Fig. \ref{4simplexfp}. This boundary has 
the equation $u =u_c$, and the value $u_c$ does not depend on $\omega$ or 
$t$.

The points on the boundary between the DC and the AS phases are found to
fall in the basin of attraction of two different fixed points:  $FP_1$ =
$(A^* = S^* = C^* =0, B^* = E^* = 0.3568)$ and $FP_2$= $(A^*= C^* =E^* =0,
B^*= 0.3568, S^*=0.61803)$. Correspondingly, we see two different
behaviors of various quantities as we cross the AS to DC phase boundary.

The first fixed point $FP_1$ is reached for all points on the AS-DC
surface with $t$ greater than a critical value $t^*$, and for large enough
$u$ even for $t < t^*$. This has two relevant eigenvalues $\lambda_1 =4$,
and $\lambda_2 =3$.

The fixed point $FP_2$ corresponds to a coexistence between the adsorbed
$SAW$ and the free collapsed globule phase. In Fig 18(a),
the line corresponding to this fixed point is shown by dotted line.
This point is reached if $t < t^*$, and $u$ is greater than, but near 
$u_c$.

The line  where the three phases meet  is also an invariant manifold for 
the renormalization group flows. We find three fixed points on this line:
\begin{enumerate}    
\item 
For $t< t^*$(= 0.34115...) the fixed point $( A^*, B^*, S^*, C^*, E^*)$ $=$ 
($\frac{1}{3}$, $\frac{1}{3}$, $0.4477$, $0.4528$, $0.0815$) is reached. The linearized equations have three
repulsive directions with eigenvalues $\lambda_{SM}= 2.2715, 
\lambda_{1}= 3.7037$ and $\lambda_2= 2.2222$. The values $\lambda_1$ and $\lambda_2$
are the same as found for the bulk $\theta$ point (see section 6.2).
\item 
For $t> t^*$ the fixed point ($\frac{1}{3}, \frac{1}{3}, 0, 0, 0.3693$) is found.
Again we find three eigenvalues greater than one, where $\lambda_{SM}= 3$ and the other
two $\lambda_1$ and $\lambda_2$ are the same as those given above.
\item 
For $t= t^*$ we find the symmetric "disordered and collapsed" fixed point ($\frac{1}{3}
, \frac{1}{3} ,\frac{1}{3}, \frac{1}{3}, \frac{1}{3}$). This fixed point has four 
eigenvalues greater than one. These values are $\lambda_{SM_1}= 2.7620$ and 
$\lambda_{SM_2}= 1.4964...$ and the other two are $\lambda_1$ and $\lambda_2$ as 
given above. 
\end{enumerate}    

The first two of these  are tetracritical points,  and the third point is
an even higher order multicritical  point ( pentacritical).

We now discuss the behavior of phase boundaries. When $t < t^* = 0.34115$, 
the critical  line $\omega=
\omega_c (u,t)$ which is the phase boundary between the  DS  and AS 
phases,  is almost linear with positive slope. Beyond  the multicritical
 point, slope of  the line separating the AS and the 
DC phases rises rather sharply. In a region
specified by $u_c < u < u_{c_1}$ (the value of $u_{c_1}$
depends on $t$) we have the coexistence between the adsorbed SAW
and the collapsed globule phase. This region is shown in 
Fig 18 by a dotted line. For $u > u_{c_1}(t)$ the line
$\omega=\omega^*(u,t)$ becomes almost flat. The value of
$u_{c_1}(t)$ decreases as $t$ is increased and becomes equal to that
of $u_{c}$ at $t=t^* = 0.34115$. At $t=t^*$ the multicritical
point becomes a symmetric ``desorbed and collapsed" pentacritical point
having four eigenvalues greater than one. 

For $t>t^*$ the  critical line $\omega = \omega^*(u,t)$ has a different
shape than for $t < t^*$. The line appears to have a maximum at
$u \leq u_{c}$. It drops rather sharply (see Fig. \ref{fig92}(b) for 
t=0.5)
in contrast to the case of $t < t^*$ at the multicritical point.
Furthermore,  the line $\omega =
\omega^*(u,t)$ for $u > u_{c}$ separating the bulk collapsed
and adsorbed phases shows the decreasing tendency as $u$ is
increased. The two tetracritical lines on the three-dimensional  $u-\omega 
- t$ phase space
meet at a pentacritical point [58]. Along one of these tetracritical 
lines, the adsorbed
(swollen) polymer coexists with both the desorbed polymer and
the desorbed globule, and the other line is the common boundary  of three 
critical surfaces of a continuous transition.

The behavior of the special adsorption line $\omega = \omega^*
(u,t)$ described above can be understood from contribution of
different coexisting polymer configurations to the bulk and
surface free energies. When both the adsorbed and desorbed
phases are in swollen state, the adsorption line has same nature
in the $\omega - u$ plane for all values of $t$, although the slope
of the line decreases as $t$ is increased. At $t = 1$, the
adsorption takes place at $\omega = 1$ and the adsorption line
in the $\omega - u$ plane has a zero slope. This is due to the
fact that at $t = 1$ and $\omega = 1$ the surface is just a part
of the bulk lattice. As $t$ is increased, $\omega$ has to be
increased to have adsorption, and since $u$ in such a situation
favors the bulk phase we have to increase the surface
attraction to counteract this tendency. In the other extreme,
{\it i.e.} when $u >> u_{c}$, the adsorption line has a
zero slope. Here the coexisting polymer configurations are those
given by $B$ and $E$ in Fig. \ref{fig91}. The free energies due to these
two configurations balance each other at all $u$ values and
therefore the line remains insensitive to the value of $u$. It
is only in the neighborhood of the special $\theta$-point that
the line becomes sensitive to the value of $t$ and $u$.

When $t < t^*$, the surface layer is strongly repulsive and
prohibits the occurrence of the $E$ configurations in the
neighborhood of the $\theta$-point. The adsorbed state is still
given by the configuration $S$, although the bulk is in the
globular compact phase. Thus to balance the free energy $\omega$
has to be increased. However, at $t > t^*$ the surface is only
moderately repulsive and therefore at certain value of $\omega$
the polymer configuration given by $E$ is formed. Thus a lower
value of $\omega$ is needed to balance the bulk free energy at
the special $\theta$-point.

A casual look at Fig. \ref{fig92}(b) may give the impression of the
existence of a re-entrant adsorbed phase as $u$ is increased.
One should, however, realize that these figures are merely a
projection on the $\omega$-$u$ plane of three dimensional
figures in which the third dimension is given by $t$. The
value of crossover exponent $\phi$ for the 4-simplex lattice is
0.7481 equal to the value found for $HB(2,3)$. 

\subsection{The 5-simplex lattice} 

\begin{figure}[ht] 
\begin{center}
\includegraphics[width=5.0in,height=2.5in]{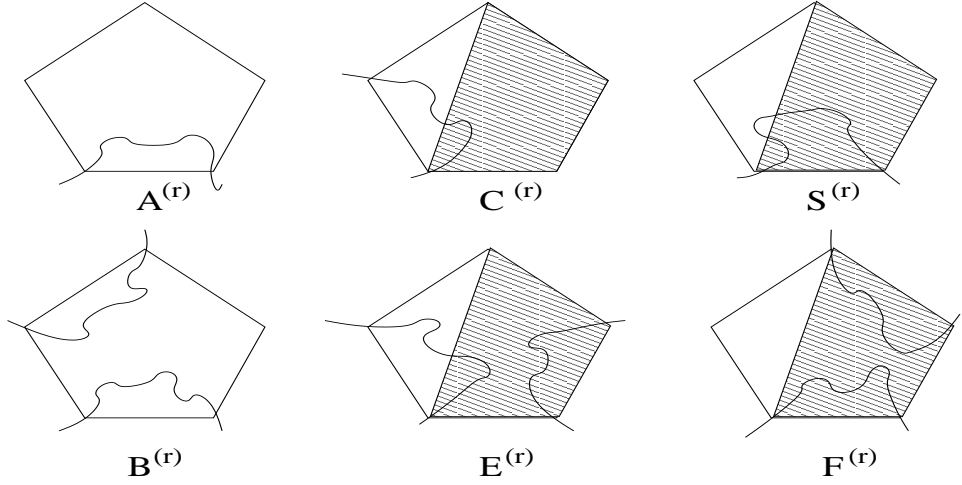}
\caption{Restricted partition functions
for the $r$th order $5$-simplex. Here $A^{(r)}$ and
$B^{(r)}$ represent the bulk generating functions for the polymer chain and
$S^{(r)}$, $C^{(r)}$, $E^{(r)}$ and $F^{(r)}$ represent the surface
functions.} 
\label{fig93}
\end{center}
\end{figure}
\begin{figure}[ht]
\begin{center}
\includegraphics[width=12cm, height=7cm]{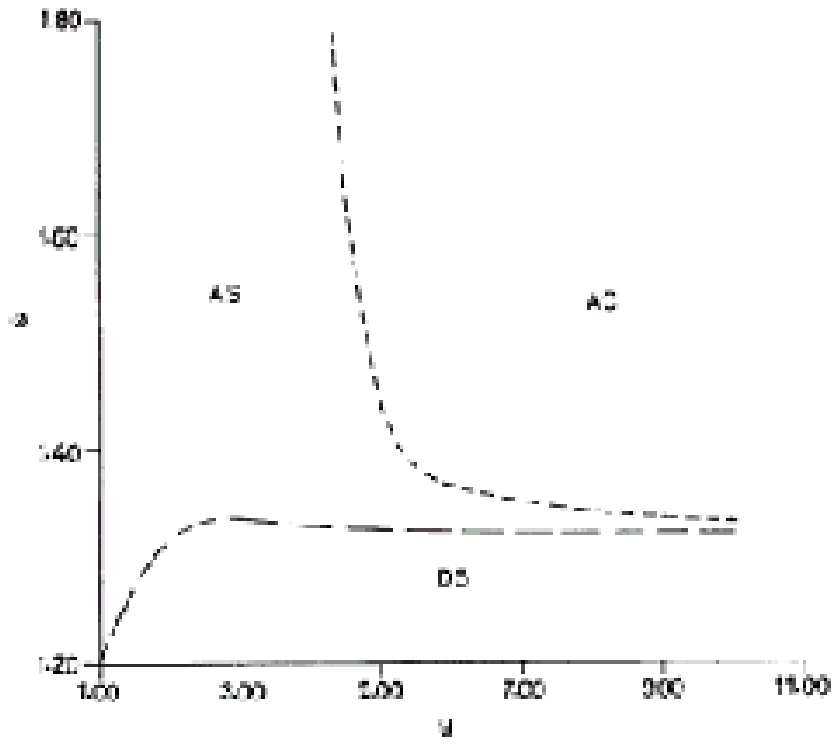}
\caption{A section of the $\omega-u-t$ phase diagram for the $5$-simplex 
at a typical value of $t =0.5$.   For other values of $t$, the phase 
boundaries shift, but the qualitative behavior is the same.}
\label{omegau}
\end{center}
\end{figure}

The grand partition function of Eq. (\ref{eq85}) for 5- simplex lattice is 
written
in terms of six restricted partition functions shown in Fig 19. Out of six
configurations two ($A^{(r)}$ and $B^{(r)}$) represent the sums of weights 
of configurations of the polymer chain within one r-th order subgraph
away from surface, and the remaining four ($C^{(r)}$, $S^{(r)}$, $E^{(r)}$ 
and $F^{(r)}$) represent the surface functions. As in the case of the 
$4$-simplex,  the recursion relations
for $A^{(r)}$ and $B^{(r)}$ do not include other variables. 

In this case, 
we have three phases: The adsorbed swollen (AS), the adsorbed collapsed 
(AC), and the desorbed swollen (DS).  It is straight forward to write down 
the fixed points corresponding to these phases from the known fixed points 
for SAWs on the $4$- and $5$- simplexs. The basins of attraction of these 
fixed points are shown in Fig. \ref{omegau}.

For the DS phase, we have $A^*, B^*$ taking the value for the swollen
phase of the $5$-simplex ( section 6.2), with $ C^*, S^*, E^*$ and $F^*$
zero. In the AS and AC phases, we have $A^*, B^*, C^*$ and $E^*$ zero, and
$S^*, F^*$ has the value of corresponding the fixed points S and C  
respectively  in 
Fig. \ref{4simplexfp} ($A^*$, $B^*$ in the terminology of section 6.2).  

The critical surface separating the DS and AS phases is a two-dimensional
surface in the $3$-dimensional parameter space $(\omega, u,t)$. All points
on this surface, under renormalization, flow to the ``symmetrical" fixed
point $(S^* = C^* = A^*, E^* = F^* = B^*)$, with values of $A^*,B^*$ same
as for the DS fixed point.  The linearization of recursion relations about
this fixed point gives two eigenvalues $\lambda_1= 3.1319$ and $\lambda_2=
2.5858$ greater than one. The line $\omega_c(u, t)$ is therefore a
tricritical line. The crossover exponent $\phi= 0.8321$.

The AS and AC phases are separated by the critical $\theta$-surface
$\omega = \omega_c(u,t)$ (see Fig 20).  For $\omega \gg {\omega_c}$(u, t),
this surface tends to the surface $u= u_c= 3.316074$ in agreement with the
critical value of $u$ for the collapse transition in the bulk of the
4-simplex lattice. However, in this case, this $\theta$-surface never
meets the AS-DS  phase boundary.  It is shown in [58] that at
$\omega \geq \omega_c (u, t)$ the $\theta$-line bends and approaches very
slowly the special adsorption line $\omega_c (u, t)$ as the value of $u$
is increased. Even for $u= 200$ the two lines have not merged.   The
whole AS-AC boudary corresponds to the basin of attraction of a single 
fixed point,  and  the parameter $t$ has no qualitative effect on the 
phase diagram.

\subsection{Surface adsorption of a SAW in  dimensions $2-\epsilon$}

We consider the family of $GM_b$ fractals and evaluate the value of the 
exponent $\phi$ and examine its behavior as $b$ is varied from 2 to $\infty$.
The energies $E_s$ and $E_t$ are defined as before. We put $E_u=0$ for 
simplicity. 

A walk is called a surface walk (and assigned configuration $S$) when it 
enters through one corner of the surface and leaves from the other. A bulk
walk represented by $B$ has no step on the surface or on the bonds 
connecting
the surface with the bulk. A walk which enters through one corner of the
surface and ends up in the bulk is assigned configuration $C$. The
generating functions for these walks can be written as
\begin{eqnarray}
 B^{(r)}(x)&=&\sum B^{(r)}x^N, \nonumber  
\\
S^{(r)}(x,\omega,t)& =&\sum S^{(r)}(N,M,R)x^N \omega^M t^R,  \nonumber \\
C^{(r)}(x,\omega,t)& = &\sum C^{(r)}(N,M,R)x^N \omega^M t^R. 
\end{eqnarray} 	 
Here $x$, as before, is the fugacity associated with each visited site
of the lattice, $B^{(r)}(x)$ is the number of distinct configurations of the 
SAW which joins two vertices of the $r$th order of the fractal in the 
bulk and $N$ is the number of sites visited by SAW, $M$ and $R$ represent 
respectively, the number of visited sites of the lattice which lie on the
surface and on a layer adjacent to the surface. Summations in Eqs.(85)
are on the repeated indices and $S^{(r)}(N,M,R)$ and $C^{(r)}(N,M,R)$ 
represent number of configurations of respective walks of the r-th order
gasket. The definition of parameters $\omega$ and $t$ is same as in sec. 
8.1.  

For $b=2$ the following three non-trivial fixed points are found:
\begin{enumerate} 
\item The fixed point $(B^*=0.61803, S^* = C^*=0)$ corresponds to the 
bulk 
state with $\nu=0.7986$. For $x=x_c(\omega)$ the fixed point is reached
for all $\omega< \omega_c(t)$. Note that $\omega_c(t)$ is a function of $t$; for $t=0.5$
the value of $\omega_c(0.5)$=1.1118.
\item The fixed point $(B^* = C^* =0, S^*=1)$ is reached for all $\omega 
>\omega_c(t)$
and $x< x_c(\omega)$. This represents the adsorbed state for the polymer 
chain
with $\nu=1$, as expected for a SAW adsorbed on a line.    
\item The fixed point $(B^*=S^*=C^*=0.61803)$ is 
obtained for $\omega=\omega_c(t)$. The linearization of Eqs.(85) about this fixed 
point yields two eigenvalues greater than one, i.e. $\lambda_s=1.6709$
and $\lambda_b=2.3819$.  
We identify this as a tricritical point. The equality of $B^*$, 
$S^*$ here shows that at this point the attraction at the surface, and 
repulsion at the next layer exactly compensate each other.  The 
crossover exponent $\phi=\frac {\log \lambda_s}{\log \lambda_b}= 0.5915$ 
At the point of adsorption transition the leading singular 
behavior of free-energy density is given as   
$f(T) \sim (T_c-T)^{2-\alpha}$ 
Thus the "specific heat" exponent $\alpha$ at the tricritical point 
is related to $\phi$ as
$\alpha=2-\frac{1}{\phi}$ 
for $b=2$ the value of $\alpha=0.3094$.
\end{enumerate} 

It is straightforward to extend this method to other members of this
family. However, as the value of $b$ increases the number of possible 
configurations of different walks increase rapidly. For $2\leq b \leq 9$  
[53,54], the value of 
$\phi$ found are 0.5915, 0.5573, 0.5305, 0.5089, 0.4908, 0.4753, 0.4617 and 
0.4497. The value of $\phi$      
decreases as $b$ increases and becomes lower than the Euclidean value
of $1/2$ at $b=6$. Zivic et al [55] have used Monte Carlo 
renormalization-group (MCRG) method to obtain the value of $\phi$ for
$2 \leq b \leq 100$ and found that lower bound of Eq(84) is violated
for $b=12$.

The limit $b\to \infty$ was analyzed by Kumar et al [53] using the finite
size scaling theory (see sec. 5) and it was found that in this limit          
\begin{equation}
\phi(b)= 1+\nu(b)(1-D_b)\left[1-\frac{(2\nu(b)-1)}{2}\frac{\log \log b}{\log b}+ \;
{\rm terms\; of \; order} \;\frac{1}{\log b} \right].
\end{equation}
As for $b\to \infty$  $\nu(b)= \frac{3}{4}$, we get $\phi(\infty)= \frac{1}{4}$. 
The first correction to $\phi$ term to finite $b$ is proportional to 
$2-{\tilde d}_b$, similar to that found for $\gamma(b)$ in Eq(71). Since
$\frac{2\nu(b)-1}{2}$ is positive for $\nu= \frac{3}{4}$, the first 
correction term in Eq(86) is negative which, when multiplied by $(1-D_b)\nu(b)$
makes a positive contribution to $\phi(b)$. This implies that $\phi(b)$
approaches to $\frac{1}{4}$ value monotonically as $b$ is increased.

Similar to the behavior of exponent $\gamma$ for SAWs discussed in section 5, 
we find $\phi(b)$ does not converge to the Euclidean value in the limit
$b \to \infty$. It has, however, been noted in [53] that the adsorbing
surface of a fractal container is similar to that of a penetrable 
surface of a regular lattice in which case $\phi=1-\nu$ should be 
satisfied [56, 57].          
    
\section{INTERACTING WALKS}

So far we were concerned with a single linear flexible polymer chain and studied
its conformational properties in different environments. We now show how the
critical behavior of two interacting long flexible linear polymer chains can
also be studied using a lattice model of two interacting walks.

Depending on the solvent quality and the attractive interactions between intrachain
and interchain monomers a system of two interacting long flexible polymer chains can
acquire different configurations. The chain may be in a state of interpenetration in
which the chains intermingle in such a way that they cannot be distinguished from
each other. Or, the two chains may get zipped together in such a way that they lie side
by side as in a double stranded DNA. It may also be possible, particularly at high
temperatures, that the two chains get separated from each other without any overlap.

By varying the temperature or tuning the interaction the system can be
transformed from one state to another. The point at which the zipped-unzipped transition
takes place is a tricritical point and in its proximity a crossover is
observed. In the asymptotic limit the mean number of monomers $M$ in
contact with each other at the tricritical point is assumed to behave as

\begin{equation} 	 
M \propto N^y,
\end{equation}
where $N$ is the total number of monomers in a chain and $y$ is the contact exponent.

A lattice model of two interacting walks on the  $n$-simplex lattices has been 
developed in \cite{kumar4,kumar5,kumar6}. 
Two different situations were considered: In one,  
the two walks have no mutual exclusion,  and a lattice bond can be 
occupied
by one step of each walks. This model has been referred to as a model of two interacting
crossed walks (TICWs). In the second model, there is mutual exclusion 
between chains 
and a lattice bond can at most be occupied by a step of either walks. This model is
called a model of two interacting walks (TIWs). In a general model of two interacting
walks one can associate an energy $E_b$ for each bond occupied by a step of both walks.
The model of TICWs corresponds to $E_b=0$ and that of TIWs to $E_b=\infty$.

In the TIWs the two chains cannot be at the same site because of mutual exclusion,
but there is lowering of energy if they occupy nearest neighbour sites. The strength
of both the inter- and intra-chain monomer interactions depend on the solvent 
and chemical nature of the monomers. Let the two chains or walks be referred to as
$P_1$ and $P_2$.
The generating functions for this system can be written as
\begin{equation} 	 
G(x_1,x_2;u_1,u_2,u_3) = \sum_{\rm all \; walks} x_1^{N_1}u_1^{R_1}
x_2^{N_2}u_2^{R_2}u_3^{R_3},
\end{equation}
where $N_1,x_1,u_1$ and $R_1(N_2,x_2,u_2 \;{\rm and} \; R_2)$ refer,
respectively, to  the number of steps, fugacity weight attached to each 
step,
the Boltzmann factor associated with the attractive interaction between
monomers and the number of pairs of nearest neighbors in the chain   
$P_1(P_2)$. $R_3$ is the number of pairs
of monomers of different walks occupying the nearest neighbor lattice
sites and $u_3$ denotes
the Boltzmann factor associated with the attractive interaction
between monomers of $P_1$ and $P_2$.

Since the individual chain can be either swollen, collapsed or at the
tricritical ($\theta$) point the variables ($x_1,u_1$) and ($x_2,u_2$) can
be taken to be known (see section 6.2). 
Therefore, Eq.(89) has only $u_3$ as independent
variable. With these simplifications it has become possible to evaluate
the restricted generating functions for the  $n$-simplex
lattices.

From the generating function of Eq.(89) one can calculate the average number of monomers
of the two chains which are in contact (nearest neighbor) with each other from the
relation  
\begin{equation} 	 
\langle R_3 \rangle = u_3\frac{\partial \log G}{\partial u_3}.
\end{equation}
The RSRG transformation has been used in [60, 61] for $n$-simplex lattices to solve 
the model exactly and calculate the phase diagram and the value of $y$ for the different 
conditions of the solvent.

Since the topological structure of a $3$-simplex lattice is such that it can
not have two SAWs on it, the model of TIWs was solved on the $4$,$5$ and $6$ simplex
lattices. These lattices can have two interacting SAWs, with possible 
self-attraction also, without
the walks crossing each other at any lattice point. On the 3- and 4- simplex
lattices TICWs model in a condition in which both chains are in the swollen state
has been solved in [61].

\begin{figure}[ht]
\begin{center}
\includegraphics[width=3.0in,height=2.5in]{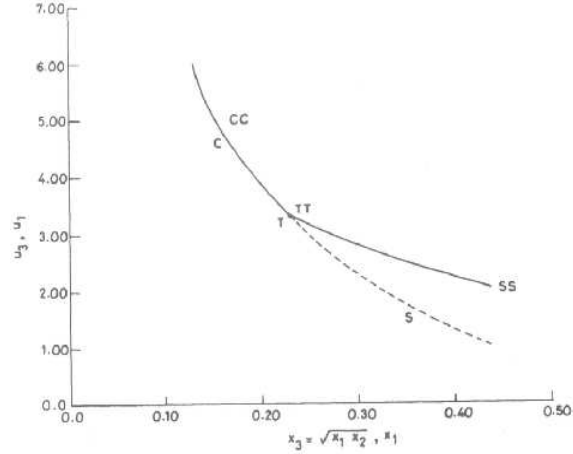}
\caption{The state of a system of two polymer chains in a non-selective
solvent on the 4-simplex lattice in the $x_3({=\sqrt {x_1 x_2}}), u_3$ plane. 
Lines $SS$ and $CC$ represent
the tricritical lines of the zipped state of two chains each in the swollen state
and the interpenetration state of the chains in a compact globule phase, respectively.
Point $TT$, at which these lines meet, represents a transition point  from a segregated to an
interpenetrated state of the chains each at its $\theta$ point. For
a given value of $x_3$ which corresponds to the swollen state of both 
chains, the chains are in interpenetrated state when the value 
of $u_3$ is less than the value given by the line $SS$. For the value of
$x_3$ corresponding to the chains in their compact globule state or at their
$\theta$ points, the two chains are in the segregated state for all values of $u_3$ less than
the value given by the line $CC$ or point $TT$. We also show the $x-u$ phase diagram
of a single chain for comparison's sake. Note that the line $CC$ overlaps with line $C,TT$
with point $T$.}
\end{center}
\end{figure}

\begin{figure}[ht]
\begin{center}
\includegraphics[width=3.0in,height=2.5in]{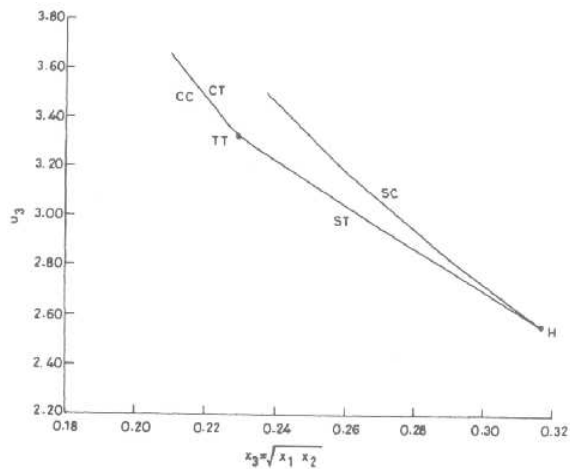}
\caption{The phase diagram of a system of two chains in a selective solvent on the 
4-simplex lattice in the $x_3(={\sqrt {x_1 x_2}}), u_3$ plane. Lines $SC$ and $ST$
respectively, represent the interpenetration states of two chains when one chain is in
a swollen state and other is in compact globule phase or at its $\theta$ point. Line
$CT$ corresponds to the configuration of interpenetration of the chains when one 
in the compact globule phase and the other is at its $\theta-$point. When the
value of $u_3$ is less than that given by corresponding lines, the two chains are
segregated from each other.} 
\end{center}
\end{figure}

\begin{figure}[ht]
\begin{center}
\includegraphics[width=3.0in,height=2.5in]{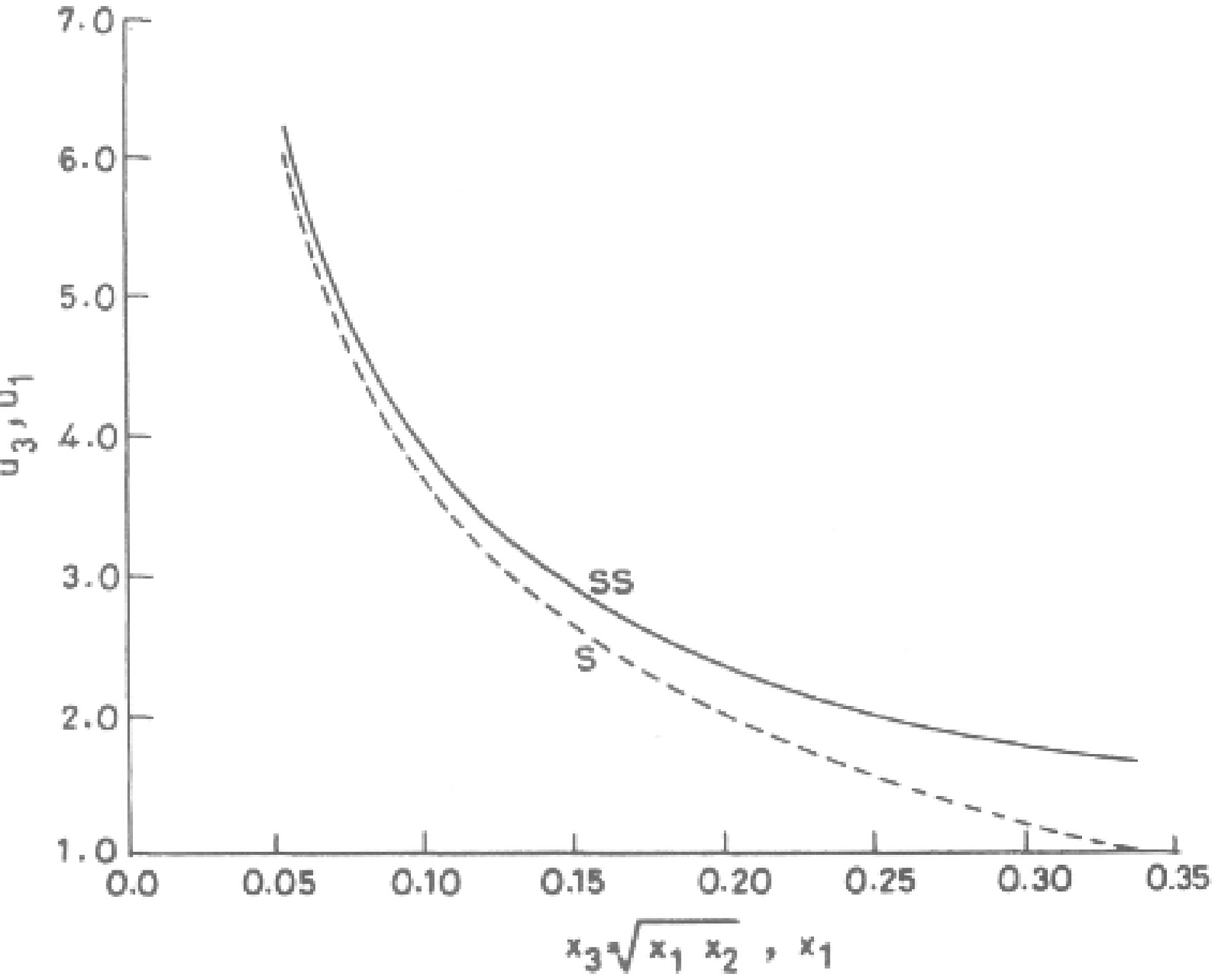}
\caption{The phase diagram representing the configurations of a system of two
chains on the 5-simplex lattice in the $x_3(={\sqrt{x_1 x_2}}), u_3$ plane. 
Line $SS$, as in Fig 22, represents the 
zipped state
of the chains. For all values of $u_3$ less than the value given by line $SS$, the two
chains intermingle with each other. Line $S$ represents the swollen state of a chain.}
\end{center}
\end{figure}
  
\subsection{TIW's on the 4-simplex lattice}

To describe the two walks on the 4-simplex lattice we need five restricted
partition functions. Two of these, $A^{(r)}$ and $B^{(r)}$ defined in Fig 11
correspond to the walk $P_1$ and identical functions $C^{(r)}$ and $D^{(r)}$ 
correspond to walk $P_2$. The restricted partition function $E^{(r)}$ represents
the configurations where walks $P_1$ and $P_2$ occupy neighboring sites and is
sum of weights of configurations in which walks $P_1$ and $P_2$ enter and exit
the r-th order subgraph once each. The two corner vertices of the subgraph are 
occupied by the walk $P_1$ and the other two corner vertices by walk $P_2$.
The recursion relation for the generating function $E^{(r)}$ involves both
$A^{(r)}$, $B^{(r)}$ and $C^{(r)}$, $D^{(r)}$ and is written as[60],
\begin{equation}   
E^{(r+1)}= A^2C^2+2ACE(A+C)+2E^4+6E^2(B^2+D^2)+4E^3(B+D).
\end{equation}   
As shown in sec. 6.2, a polymer chain on a 4-simplex lattice can be in any 
one of the states of the swollen, compact globule and the $\theta$-point and
is described in the asymptotic limit by the fixed points (0.4294.., 0.04998..),
(0, $22^{-\frac{1}{3}}$) and ($\frac{1}{3}, \frac{1}{3}$),  
respectively. The fixed points
corresponding to a swollen state is reached for all values of $u_1$ (or
$u_2$) $<u_c$=3.31607 at $x_1$ (or $x_2$)=$x_c$. The value of $x_c$ is
a function of the interaction $u_1$ (or $u_2$). The end to end 
distance for a chain of N-monomers in this state varies as $N^{\nu}$ with 
$\nu$=0.7294... . The fixed point corresponding to the compact globule
state is reached for all values of $u_1$ (or $u_2$) $>u_c$ (i.e. at low
temperatures) at $x_1$(or $x_2$)=$x_{c}(u_i)$, where i=1 or 2. In a 
compact globule state the polymer chain has a finite density of the monomer
per site when N$\rightarrow \infty$. At $u_1$(or $u_2$)=$u_c$=3.31607 and 
$x_1$(or $x_2$)=$x_c$($u_c$)=0.22913... the $\theta$-point state is reached.

In a system of two chemically different polymer chains, we may have six
independent combinations of the individual chains which we indicate by
$SS, CC, TT, SC, ST$ and $CT$, where letters $S, C$ and $T$ now stand for 
the swollen, compact globule and $\theta$-point states. The recursion relation
(91) is solved using the fixed points ($A^*, B^*$) and ($C^*, D^*$) corresponding 
to the given states of the chain $P_1$ and $P_2$ and the starting weight 
$E^{(1)}= (x_1x_2)^2u_{3}^4$. The solution leads to two fixed 
points denoted as $E_{s}^{*}$ and $E_{i}^{*}$ for each combination of
states of the individual chains. In table I the values of these fixed 
points, relevant eigenvalues and the contact exponents $y$ at the tricritical
points are listed \cite{kumar5}. Figs. 19 and 20 show the state of two chains 
in the $x_3(={\sqrt {x_1 x_2}}), u_3$ plane respectively, for non-selective and 
selective solvents.

Lines $SS$ and $CC$ in Fig. 19 represent the tricritical lines of 
the zipped states of two chains each in the swollen state and the 
interpenetrating state of the chains each in a compact globule phase, 
respectively. Point $TT$, at which these lines meet, represents a
transition point from a segregated to an interpenetrated state of the
chains each at its $\theta$ point. For a given value of $x_3$ which 
corresponds to a swollen state of both chains, the chains are in the 
interpenetrated state when the value of $u_3$ is less than the
value given by the line $SS$. For the values of $x_3$, corresponding
to the chains in their compact globule state or at their $\theta$-point,
the two chains are in a segregated state for all values of $u_3$ less 
than the values given by line $CC$ or point $TT$. The $x,u$ phase
diagram of a single chain is also given in the figure for the 
comparison sake.

The state of two interacting chains in a selective 
solvent on the $4$-simplex lattice is shown in Fig 20. Lines
$SC$ and $ST$, respectively, represent the interpenetration state of two 
chains when one is in the compact globule phase or at its $\theta$-point.
Line $CT$ corresponds to the configuration of the interpenetration
of the chains when one is in the compact globule phase and the
other is at its $\theta$-point. When the value of $u_3$ is less
than given by the corresponding lines, the two chains are segregated from 
each other. The individual chains configuration remains unchanged 
whether they are segregated or intermingled.         

\subsection{\bf TIW's on the 5-simplex lattice}

For the 5-simplex lattice we also need five restricted partition functions
to describe the generating functions of two walks; two corresponding to
walk $P_1$, two corresponding to walk $P_2$ and $E^{(r)}$ which represents 
the configurations in which walks $P_1$ and $P_2$ enter and exit the r-th order
subgraph once each and may occupy neighboring sites. The recursion relation
for $E^{(r)}$ in this case is lengthy[60] and is therefore not reproduced here. 

As already shown in section 6.2 a polymer chain always remains in a swollen
state for all values of interaction, (self-) attraction on a 5-simplex lattice.
This state is characterized by the fixed point (0.3265..., 0.0279...).
In this case we therefore have only one combination, i.e. $SS$ of chains.
Using these values of fixed points for single chains, the recursion relation
for the partition function $E^{(r)}$ of two interacting chains (see \cite{kumar5})
is solved and fixed points $E_s^{*}=0.0279$ or $E_i^{*}=0.2713$ are found. The
fixed point $E_{s}^{*}$ is found for all values of $u_3 < u_{3c}(x_3)$.
At $u_3=u_{3c}(x_3)$ fixed point $E_{i}^{*}$ is found.           


\begin{landscape}

\begin{table}
\label{table1}
\centering
\caption{Values of fixed points, relevant eigenvalues, and the contact exponent $y$ at the
critical point for the 4-simplex lattice. The swollen, compact globule, and the tricritical
configuration of each chain is indicated by letters $S,C,\; {\rm and}\; T$ respectively.
$\lambda_1$ is the largest eigenvalue of the system, and $\lambda_i$ refers to the two
chain configuration point.}
\begin{tabular}{llllll}  \\ \hline 
State \\ of individual chains & $E^*_s$ & $E^*_i$ & $\lambda_i$ & $\lambda_1$ & $y$ \\
\hline 
SS & 0.04998 & 0.6125      & 2.6420 & 2.7965 & 0.9447 \\
CC & 0.0     & 22$^{-1/3}$ & 2.5440 & 4.0    & 0.6735 \\
TT & 0.01484 & 1/3         & 2.2222 & 3.7037 & 0.6098 \\
SC & 0.0     & 0.4880      & 2.8520 & 4.0    & 0.7559 \\
ST & 0.02687 & 0.4294      & 2.8526 & 3.7037 & 0.6311 \\
TC & 0.0     & 0.3680      & 2.5740 & 4.0    & 0.6820 \\ 
\hline 
\end{tabular}
\end{table}
\begin{table}
\label{table2}
\centering
\caption{Values of fixed points, $G_s^*,H_s^*,I_s^*,G_I^*,H_I^*$ and $I_I^*$. The 
labeling is the same as in Table 1, but for 6-simplex.}
\begin{tabular}{llllllllll}  \\ \hline 
State & $G^*_s$ & $H^*_s$ & $I^*_s$ & $G^*_i$ & $H^*_i$ & $I^*_i$ & $\lambda_i$ & 
$\lambda_1$ & $y$ \\
of ind. & & & & & & & & & \\ 
chains & & & & & & & & \\ \hline 
SS & 0.0175  & 0.0007    & 0.0007 & 0.1406 & 0.0147 & 0.0147 & 2.2054 & 3.4965 & 0.6318 \\
CC & 0.0     & 0.0       & 0.0    & 0.0    & 0.0713 & 0.0713 & 4.2201 & 0.6318 & 0.7902 \\
TT & 5.4$\times 10^{-4}$ & 5.4$\times 10^{-4}$ & 5.4$\times 10^{-4}$  & 0.0957 & 0.0535 & 
0.0535 & 3.2225 & 5.4492 & 0.6902 \\
SC & 0.0  & 0.0  & 0.0 & 6.75$\times 10^{-5}$ & 0.1049 & 0.0711 & 5.0529 & 6.0 & 0.9041\\
ST & 0.0028 & 0.0003  & 0.001 & 0.229 & 0.1047 & 0.0776 & 5.2434 & 5.4492 & 0.9659 \\
CT & 0.0     & 0.0  & 0.0 & 0.0060   & 0.0714 & 0.0806 & 4.4019 & 6.0 & 0.8271 \\ 
\hline 
\end{tabular}
\end{table}
\end{landscape}


In Fig. 21, $u_3$ is plotted as a function of $x_3=\sqrt{x_{1c}x_{2c}}$.
The interpenetrated state is found for all values of $u_3$ which lie
 below lines $SS$ at a given $x_3$. When the values of $u_3$ reaches a
value given by the line $SS$, the two chains are zipped together. Line $S$
indicates the critical values of fugacity $x_c$ and the self-attraction
$u_1$ (or $u_2$) of a polymer chain.    

\subsection{\bf TIW's on the 6-simplex lattice}

In this case one needs nine generating functions[60]; three corresponding 
to chain $P_1$, three corresponding to chain $P_2$ and the remaining
 three denoted as $G^{(r)}$, $H^{(r)}$, and $I^{(r)}$ 
represent, respectively, the sum of weights of configurations in 
which walks $P_1$ and $P_2$ enter and exit the r-th order subgraph once
each, $P_1$ twice and $P_2$ once, and $P_1$ once and $P_2$ twice.  
Since the 6-simplex lattice exhibits the collapse transition there are six
independent combinations of single chain states, similar to the case
of 4-simplex lattice. The fixed points corresponding to single chains
found in section 6.2  have been used to find the fixed points of the 
recursion relation for  the restricted partition functions for $G, H,$ and $I$
with suitable starting weights \cite{kumar5}. The values of fixed points , 
corresponding eigenvalues and contact exponents are given in table II.
The qualitative features of the phase diagram found for this lattice
are the same as that of the 4-simplex lattice.

Though the model of TIW's discussed above ignores the effects of one chain
on the critical behavior of the other chains, it provides a qualitative 
description of the phase diagram of systems of two polymer chains in a 
solution which may have different qualities for different chains and 
may serve as a starting point for more thorough investigation of
segregation and entanglements in a real systems. The phase diagrams 
plotted here are in the plane $x_3(={\sqrt {x_{1} x_{2}}}$, $u_3$. 
Plot of phase diagrams in a three-dimensional $u_1, u_2, u_3$ space
is expected to give more informations about the states the two interacting
chains.

\section{CONCLUDING REMARKS}

  It seems  fair to say that the study of linear and branched polymers on 
fractals has been very useful in developing a better understanding of 
the critical behavior of polymers. One can verify the general qualitative
features of the polymers on fractals which are very often similar to that in
real experimental systems. One has a good deal of freedom in selecting the 
details of the fractal, and this can be used to find one that represents 
best the local geometry of the space. The exact values of the critical 
exponents do depend on the details of the fractal.  But what is more 
important is that one can handle the different interactions in the 
problem, between different monomers, with the substrate, or with a 
different chain consistently and satisfactorily in way that allows
exact calculation.

  Of course there are many unsolved problems, and possible directions for 
further research in this area. The most interesting problem would be to 
try to extend these exact solutions  to some  fractals with infinite 
ramification index. There are some studies of statistical physics models 
of interacting degrees of freedom on Sierpinski carpets, using Monte Carlo
simulations, or approximate renormalization group using bond-moving, or 
other ad-hoc approximations. An exactly soluble case would be very 
instructive here. 
 
 Determination of the exact scaling functions, not just the critical
exponents, for these problems would be instructive: For example, the exact
functional form of the scaling function of the probability distribution of
the end-to-end distance of the SAW's on these fractal, or the periodic
function of Fig. \ref{logperiodic}. For a numerical study of the former,
see \cite{ordemann}. Solving the SAW problem with quenched disorder is
another interesting question. For the $3$-simplex fractal, this
corresponds to making the variables $B^{(r)}$ random variables, and one
has to determine the probability distribution of this variable for large
$r$.

 We have discussed only the equilibrium properties of polymers. Of course,
in many real systems, the time scales for equilibriation can be very 
large. It is thus of interest to study non-equilibrium properties of 
statistical mechanical systems on fractals. A simple prototype is    
the study of kinetic Ising model on  fractals. Closer to our interests 
here, one can study, say, the  reptation motion of a polymer on the fractal 
substrate. This seems to be a rather good first  model of motion of a 
polymer in gels. 
  
 It is a pleasure to thank Sumedha for a careful reading of the 
manuscript, and her many constructive suggestions for an  improved 
presentation.



\end{document}